\documentclass[fleqn,usenatbib]{mnras}

\usepackage{newtxtext,newtxmath}

\usepackage[T1]{fontenc}

\DeclareRobustCommand{\VAN}[3]{#2}
\let\VANthebibliography\thebibliography
\def\thebibliography{\DeclareRobustCommand{\VAN}[3]{##3}\VANthebibliography}

\usepackage{graphicx}	
\usepackage{amsmath}	
\usepackage{comment}
\usepackage[dvipsnames]{xcolor}

\graphicspath{{./picture/}}

\newcommand{\Msun}{\mathrm{M}_{\odot}} 
\newcommand{\cmc}{\mathrm{cm}^{-3}}
\newcommand{\mdotsunyr}{\mathrm{M}_\odot \mathrm{yr}^{-1}}
\title[First cold accretion and supermassive stars]{First emergence of cold accretion and supermassive star formation \\
in the early universe}

\author[M. Kiyuna et al.]{
Masaki Kiyuna,$^{1}$\thanks{E-mail: kiyuna@tap.scphys.kyoto-u.ac.jp (KTS)}
Takashi Hosokawa,$^{1}$
and Sunmyon Chon$^{2}$
\\
$^{1}$Department of Physics, Graduate School of Science, Kyoto University, Sakyo, Kyoto 606-8502, Japan\\
$^{2}$Astronomical Institute, Graduate School of Science, Tohoku University, Aoba, Sendai 980-8578, Japan
}

\date{Accepted XXX. Received YYY; in original form ZZZ}

\pubyear{2022}

\begin{document}
\label{firstpage}
\pagerange{\pageref{firstpage}--\pageref{lastpage}}
\maketitle

\begin{abstract}
We investigate the first emergence of the so-called cold accretion, the accretion flows deeply penetrating a halo, in the early universe with cosmological N-body/SPH simulations. We study the structure of the accretion flow and its evolution within small halos with $\lesssim 10^8~\Msun$ with sufficiently high spatial resolutions down to $\sim 1 \ {\rm pc}$ scale. 
While previous studies only follow the evolution for a short period after the primordial cloud collapse, we follow the long-term evolution until the cold accretion first appears, employing the sink particle method. 
We show that the cold accretion emerges when the halo mass exceeds $\sim 2.2\times 10^7 \ {\rm M}_\odot\left\{\left(1+z\right)/15 \right\}^{-3/2}$, the {\it minimum} halo masses above which the accretion flow penetrates halos. We further continue simulations to study whether the cold accretion provides the dense shock waves, which have been proposed to give birth to supermassive stars (SMSs). We find that the accretion flow eventually hits a compact disc 
near the halo centre, creating dense shocks over a wide area of the disc surface. 
The resulting post-shock gas becomes dense and hot enough 
with its mass comparable to the Jeans mass $M_{\rm J}\sim 10^{4-5} \ {\rm M}_\odot$, 
a sufficient amount to induce the gravitational collapse, leading to the SMS formation. 
\end{abstract}



\section{INTRODUCTION}

More than 200 quasars have been observed at $z \gtrsim 6$ \citep{Mortlock+2011, Banados+2018, Matsuoka+2019, Wang+2021},
indicating that Supermassive Black Holes (SMBHs) with $M_{\rm BH}=10^{8-10} \ \Msun$ already exist in the early universe.
The most distant quasar observed so far is located at $z=7.54$ \citep{Banados+2018},
which corresponds to the cosmic age of $0.7~$Gyr.
Theoretically, it is challenging to form SMBHs in such an early universe \citep{Inayoshi2020}.
The BHs provided by Population (Pop) I\hspace{-1pt}I\hspace{-1pt}I stars are one of the possible candidates that will grow into the observed high-z SMBHs.
Recent numerical simulations have shown that the Pop I\hspace{-1pt}I\hspace{-1pt}I stars appear with masses of $M=10$--$10^3 \ \Msun$ at $z \simeq 20$-$ 30$ \citep{HosokawaYork2012, Hirano+2014, Hirano2015, SusaHasegawa+2014, Hosokawa+2016, StacyBromm+2016, Sugimura2020},
some of which finally collapse into BHs with negligible mass loss \citep{HegerWoosley2002, TakahashiYoshidaUmemura2018}.
To attain the observed mass of the SMBHs until $z=7.5$, the seed BH should maintain the Eddington accretion rate for the entire period of the corresponding cosmic age.
However, the Eddington rate is hard to achieve due to the feedback associated with star formation and the mass accretion onto BHs \citep{JohnsonBromm2007, Alvarez+2009, Jeon2012}.

One of the most promising scenarios to form SMBHs is the ``Direct Collapse (DC)'' scenario, in which the heavy seed BHs are provided by supermassive stars (SMSs) with masses of $M=10^{4-6}~\Msun$ \citep{BrommLoeb2003}.
They potentially form in atomic cooling halos (ACHs) with $M_{\rm halo}=10^{7-8} \ \Msun$,
inside which the cloud collapse is triggered by Ly$\alpha$ cooling \citep{Omukai2001}.
The cloud monolithically collapses into a protostellar core \citep{InayoshiTasker2014}, and the mass accretion rate onto it is $0.1$ -- $1~\mdotsunyr$ \citep{Latif+2013,Chon+2018}.
A protostar continues to grow until it becomes $10^{4-6}~\Msun$ \citep{HosokawaYork2012, HosokawaYork2013}, when it collapses into a BH equally massive to the progenitor SMS \citep{ShibataShapiro+2002,Umeda+2016,Haemmerle2018}.
This model provides heavier seed BHs than those by Pop I\hspace{-1pt}I\hspace{-1pt}I model, which is advantageous to form SMBHs in the early universe.


 A key process in forming SMSs is suppressing H$_2$ cooling during the cloud collapse. If H$_2$ cooling is completely suppressed, the gas cannot cool below $T \simeq 8000 \ {\rm K}$.
Photodissociation by the Far-Ultraviolet (FUV) radiation is one of the most promising processes to disable H$_2$ cooling.
The critical FUV intensity required for SMS formation is $J_{21}\simeq10^3$ \citep{Omukai2001,Shang2010}, 
although it depends on the spectrum of radiation \citep{Sugimura+2014},
where $J_{21}$ is the specific intensity at the Lyman Werner band $(h\nu = 11.2-13.6 \ {\rm eV})$ normalized by $10^{-21} \ {\rm erg \ cm^{-2} s^{-1} Hz^{-1} str^{-1}}$.
Such large FUV intensity is mainly provided by the nearby stars or star-forming galaxies \citep{Dijkstra+2008, Holzbauer+2012}, 
since the background intensity $J_{21}\sim 10$ is much smaller than the critical value \citep{Agarwal+2012, Johnson+2013}.
\citet{Chon+2016} have conducted the cosmological simulation and found a number of halos that satisfy large FUV intensity and
demonstrated that some of them finally form SMSs at the halo centre.
However, the DCBHs formed in the FUV scenario have difficulty in mass growth, since their forming places are separated by $1$--$10~$kpc from the star-forming galaxies with massive gas reservoirs.
They wander outskirts of the galaxy with negligible mass growth \citep{Chon2021}.
Other models are proposed to form SMSs, the turbulent motion induced by the major or minor mergers of the host halo \citep{Wise2019, Latif2022} or the baryonic streaming motion \citep{Schauer+2017, Hirano2017} can delay the star formation. In these models, the cloud collapse occurs once the halo virial mass reaches $8000~$K and the mass accretion toward the cloud centre becomes $0.1$--$1~\mdotsunyr$ to form SMSs \citep{Hirano2017,Latif2022}.

The collisional dissociation is another process to destroy H$_2$,
while it requires a high-density shock in the primordial environments \citep{InayoshiOmukai2012, Inayoshi2015, InayoshiLi2018}.
When the gas is shock-heated and satisfies the conditions $n_{\rm H} \gtrsim 10^4\ {\rm cm^{-3}}$ and $T \gtrsim 5000\ {\rm K}$, 
the dissociation rate by the collision becomes more efficient than the rate of H$_2$ formation.
This region in the density-temperature diagram is called the "Zone of No Return" (ZoNR). \citet{InayoshiTasker2014} have demonstrated that starting from the initial gas density and temperature in ZoNR,
the cloud collapse proceeds with negligible H$_2$ cooling to form the protostars.
Protogalactic collisions \citep{Mayer+2010, Inayoshi2015} and baryonic streaming motion \citep{Schauer+2017,Hirano2017} are possible processes creating such dense shocks.

\citet{InayoshiOmukai2012} primarily consider cold accretion, which delivers dense and cold gas into the halo center through the cosmic filament.
This phenomenon is observed in the numerical simulation in the context of the formation of the matured galaxy with the mass of
$M_{\rm halo}= 10^{10-14}\ \Msun$ \citep{BirnboimDekel2003,DekelBirnboim2009, Keres2005, Ocvirk+2008, Brooks+2009}.
In the classical picture of galaxy formation \citep[e.g.][]{ReesOstriker1977},
the accreted gas experiences shock at the halo surface and is heated to the virial temperature of the halo,
which is termed "hot accretion".
Once the accreting gas experiences the shock, the accretion is decelerated by the thermal pressure and
it falls inward to the halo centre with the velocity comparable to the sound speed.
When the radiative cooling is efficient in the accreting flow, the shocked gas is quickly cooled, and the kinetic energy is hardly thermalized.
This allows the accreting flow keeping super-sonic,
which penetrates deep inside the virial radius of the halo without any deceleration.
\citet{BirnboimDekel2003} have analytically derived the condition to realize the cold accretion, assuming the spherical symmetry.
They found that it realizes when the halo mass is smaller than $\sim 10^{11-12} \ \Msun$.
Cosmological hydrodynamical simulations have shown that such cold accretion mainly occurs in the cosmic filament since it has a higher density than the cosmic mean value and the cooling is more efficient in such a dense region
\citep{Keres2005, DekelBirnboim2006, DekelBirnboim2009}.

Our main purpose is to provide the condition to realize the cold accretion in the early universe and to clarify whether the SMS formation is triggered by it. Whether cold accretion is realized in the early stage of galaxy formation is uncertain.
\citet{WiseAbel2007} and \citet{Greif2008} have found that the cold flow penetrates inside the virial radius of the halo,
when the halo mass is $M_{\rm halo}=10^{7-8} \ \Msun$ at $z>10$.
In contrast, \citet{Fernandez2014} have shown that the accretion flow is virialized at the halo surface and no cold accretion is observed.
They conclude that it is difficult to shock-heat the gas to enter the ZoNR by the cold accretion,
which is expected by \citet{InayoshiOmukai2012} to form SMSs.
Although the halo studied in \citet{Fernandez2014} satisfies the condition proposed by \citet{BirnboimDekel2003}, 
they do not observe any cold mode accretion.
This indicates that there should be additional requirements to trigger the cold accretion in the early universe.
In primordial environments, the cooling rate decreases sharply below $T=10^4 \ {\rm K}$, 
which can pose a lower mass limit for the cold accretion.
We construct a semi-analytic model to describe the accretion flow and quantitatively estimate the thermal state of the shocked gas in the halo with $10^{6}~\Msun~\leq~ M_\text{halo} ~\leq 10^{10}~\Msun$ to derive the condition to realize cold accretion.
We further conduct cosmological simulations to study whether the cold accretion emerges in the halo with $M_{\rm halo}=10^{7-8} \ \Msun$. 
We discuss the possibility of the SMS formation by the shock-heating caused by the cold accretion.

The rest of the paper is organized as follows. We describe our simulation methods in Section~\ref{sec:methods}. In Section~\ref{sec:results}, we present our simulation results, paying attention to the dynamics of the accretion flow within dark halos. We show the minimum halo masses above which the cold accretion appears, which is well interpreted by a semi-analytic model. We further investigate the possible SMS formation, based on our simulation results in Section~\ref{sec:SMS_pre}. 
We finally provide discussion in Section~\ref{sec:discussion} and concluding remarks in Section~\ref{sec:conclusions}.

 \section{METHODS}
\label{sec:methods}

We perform a suite of cosmological simulations using the N-body + ${\rm SPH}$ code {\tt GADGET-3} \citep{Springel2005}.  We set up three different realizations of the initial conditions with {\tt MUSIC} \citep{HahnAbel2013} at the redshift $z=99$. The size of the cosmological volume is $V=(1 \ h^{-1} \ {\rm cMpc})^3$. We first conduct DM-only N-body simulations to follow the halo assembly histories. 
We use $(256)^3$ DM particles, whose mass is $m \simeq 4360 h^{-1} \ \Msun$. We identify the most massive halo in each case, halos A, B, and C, with the {\tt Rockstar} halo finder \citep{Behroozi2013} at the epoch of $z \sim 10$. These halos have the mass of $M_{\rm halo} \sim 10^8 \ \Msun$. Below we show that the cold accretion caused by Ly$\alpha$ cooling first emerges in these halos before $z \sim 10$.
Table \ref{table:halo_penetration} summarizes the properties of these halos at two characteristic evolutionary stages. Hereafter we use the following definition of the virial temperature as
 \begin{eqnarray}
  T_{\rm vir}&\equiv& \frac{GM_{\rm halo}\mu m_{\rm H}}{2k_{\rm B}r_{\rm vir}}\nonumber\\
  &\simeq& 2.65 \times 10^4\ {\rm K}\left(\frac{M_{\rm halo}}{10^8 h^{-1}\ {\rm M}_\odot}\right)^{2/3}
  \left(\frac{1+z}{10}\right) ,
\end{eqnarray}
where $\mu\simeq 1.3$ is the mean molecular weight, $m_{\rm H}$ is the mass of a hydrogen atom, $G$ is the gravity constant, and $k_{\rm B}$ is Boltzmann constant.
The virial radius of halo $r_{\rm vir}$ is defined as
\begin{eqnarray}
    r_{\rm vir}&=&
    \left\{\frac{GM_{\rm halo}}{9\pi^2 H_0^2 \Omega_{\rm m,0}(1+z)^3}\right\}^{1/3}
\end{eqnarray}
with its mass $M_{\rm halo}$ at the epoch of redshift $z(>2)$, where $H_0$ is the Hubble constant and $\Omega_{\rm m,0}$ is the density parameter of matter.

\begin{table}
  \caption{Physical quantities of halos A, B, and C at the two characteristic stages of (i) $T_{\rm vir} = 8000\ {\rm K}$, and (ii) the final snapshot when $M_{\rm halo}=10^8\ \Msun$.}
  \label{table:halo_penetration}
  \centering
  \begin{tabular}{l|cccc}
    \hline
    & $z$  &  $M_{\rm halo}\ [{\rm M}_\odot]$ &$T_{\rm vir}\ [{\rm K}]$
     &$r_{\rm vir}\ [{\rm kpc}]$\\
  \hline \hline
  A & $21.4$  & $5.0\times 10^{6}$ & $8.0\times 10^3$ & $0.18$ \\
    & $10.1$  & $1.0\times 10^{8}$ & $3.1\times 10^4$ & $1.01$\\
      \hline
  B & $14.7$  & $8.4\times 10^{6}$ & $8.0\times 10^3$ & $0.31$\\
    & $10.0$  & $1.0\times 10^{8}$ & $2.9\times 10^4$ & $1.02$\\
      \hline
  C & $14.5$  & $8.7\times 10^{6}$ & $8.0\times 10^3$ & $0.32$\\
    &  $8.9$  & $1.0\times 10^{8}$ & $2.6\times 10^4$ & $1.14$\\
    \hline
  \end{tabular}
\end{table}

We next perform "zoom-in" simulations considering baryonic physics.
We set the zoom-in region of $(0.4 h^{-1} \ {\rm Mpc})^3$ centred on the target halos. Starting from the same initial conditions as for the N-body simulations, we re-simulate the evolution until the target halos grow to $M_{\rm halo} = 10^8 \ \Msun$. In the zoom-in region, we homogeneously distribute $(1024)^3$ DM and gas particles effectively as the initial state of each run. The resulting mass resolutions are $68.1h^{-1}\ \Msun$ for the DM, and $11.6h^{-1}\ \Msun$ for the gas, respectively. These values are sufficiently small to resolve the structure of $\sim 10-10^2$~pc, or a gas disc forming near the halo centre. The mass resolution outside the zoom-in region is the same as in the original DM-only simulations. 
We set the cosmological parameters at the PLANCK13 values of $\Omega_{\rm m} =0.3086, \Omega_\Lambda=0.6914, \Omega_{\rm b}=0.045, h=0.6777, \sigma_8=0.8288, n_{\rm spec}=0.9611$ \citep{Planck13} for all the above simulations.


We solve the non-equilibrium chemistry network with 20 reactions among 5 species, e$^{-}$, H, H$^{+}$, H$_2$, and H$^{-}$,
with an implicit scheme \citep[see][for details]{YoshidaAbel2003,YoshidaOmukai2006}. 
We include in the energy equation the heating and cooling processes associated with chemical reactions, radiative cooling by Ly$\alpha$ and H$_2$ line emission.


Following \citet{Fernandez2014}, we assume constant Lyman-Werner (LW; $11.2-13.6 \ {\rm eV}$) background radiation field $J_{21}=10$ throughout our simulations. 
Previous studies estimate $J_{21} \sim 10$ as the cosmological mean values achieved for the period of $25 \gtrsim z \gtrsim 10$ \citep{Dijkstra+2008,Holzbauer+2012}, and it is high enough to suppress the Pop I\hspace{-1pt}I\hspace{-1pt}I star formation in mini-halos \citep{Haiman+2000}.
Note that $J_{21}$ required for the SMS formation solely by photodissociating H$_2$ molecules is $J_{21} \gtrsim 10^3$ \citep{Shang2010, Sugimura+2014}. Since we pursue an alternative SMS formation channel by H$_2$ collisional dissociation in dense shocks, we only assume the moderate value of $J_{21} = 10$. We follow the long-term evolution during which the cold accretion first emerges, allowing the normal Pop I\hspace{-1pt}I\hspace{-1pt}I star formation via H$_2$ cooling in ACHs.


We employ the sink particle method to deal with the small-scale star formation process with saving computational costs \citep{Hubber+2013, Chon+2017}. 
In the ACH, for instance, Ly$\alpha$ cooling causes the run-away collapse of a gas cloud, leading to the normal Pop I\hspace{-1pt}I\hspace{-1pt}I star formation.
\citet{BateBurkert1997} argue that the Jeans mass must be resolved by at least $\simeq 80$ particles to follow the self-gravitating gas dynamics accurately. With this criterion, we estimate the maximum density $n_{\rm H}$ resolvable in our simulation as
\begin{eqnarray}
n_{\rm H}&\simeq&\frac{1}{(80m)^2 \mu m_{\rm H}}
\left(\frac{\gamma k_{\rm B}T}{G\mu m_{\rm H}}\right)^3 \nonumber \\
&\simeq& 2.7\times 10^8\ {\rm cm^{-3}}\left(\frac{m}{11.6h^{-1}\ {\rm M}_\odot}\right)^{-2}
 \left(\frac{T}{10^4{\rm \ K}}\right)^3 , 
 \label{eq:nH_limit}
\end{eqnarray}
where $m$ is gas particle mass and $\gamma\simeq 5/3$ is the adiabatic index. This is sufficiently higher than the density threshold of ZoNR, $n_{\rm H} \sim 10^4\ {\rm cm^{-3}}$ at $T \simeq 5000 \ {\rm K}$ \citep{InayoshiOmukai2012}.
After H$_2$ cooling becomes effective, however, the gas temperature decreases down to 
$T \simeq 2000~{\rm K}$\footnote{
The gas temperature does not decrease to hundreds of Kelvin
for $n_{\rm H} \gtrsim 10^2 \ \cmc$ because of our omission of the self-shielding effect on ${\rm H_2}$ photodissociation. This treatment does not affect the large-scale gas dynamics and chemo-thermal evolution in the ZoNR, as noted in detail in Section \ref{subsubsec:dense_shock}.
}
as the density rises to $\sim 10^6 \ {\rm cm^{-3}}$. Equation \eqref{eq:nH_limit} provides the lower threshold density $n_{\rm H}\simeq 2.2 \times 10^6 \ {\rm cm^{-3}}$ for such a case. We thus insert sink particles when the gas density exceeds $n_{\rm H}= 2 \times 10^6\ {\rm cm^{-3}}$. To prevent the artificial fragmentation by doing so, we make the equation of state adiabatic for slightly lower densities, $n_{\rm H} \geq 10^6\ {\rm cm^{-3}}$. In our simulations, the temperature is normally higher than 2000~K, meaning that we resolve the Jeans mass with more than 80 gas particles.


We assume each sink particle accretes the gas particles within the radius 10 times larger than the smoothing length. The actual values of the sink radius are $\simeq 2-3 \ {\rm pc} \simeq (0.002-0.006) \ r_{\rm vir}$. We examine effects by changing the sink radius in Section \ref{subsec:sink}.
For simplicity, we ignore any feedback effects from accreting sink particles, including radiative feedback from massive stars or BHs, mechanical feedback, and metal enrichment caused by supernova explosions. We discuss these additional effects in Section~\ref{subsec:feedback}.

\section{SIMULATION RESULTS}
\label{sec:results}

\begin{figure}
	\includegraphics[bb=0 0 480 1420,width=6cm,scale=0.2]{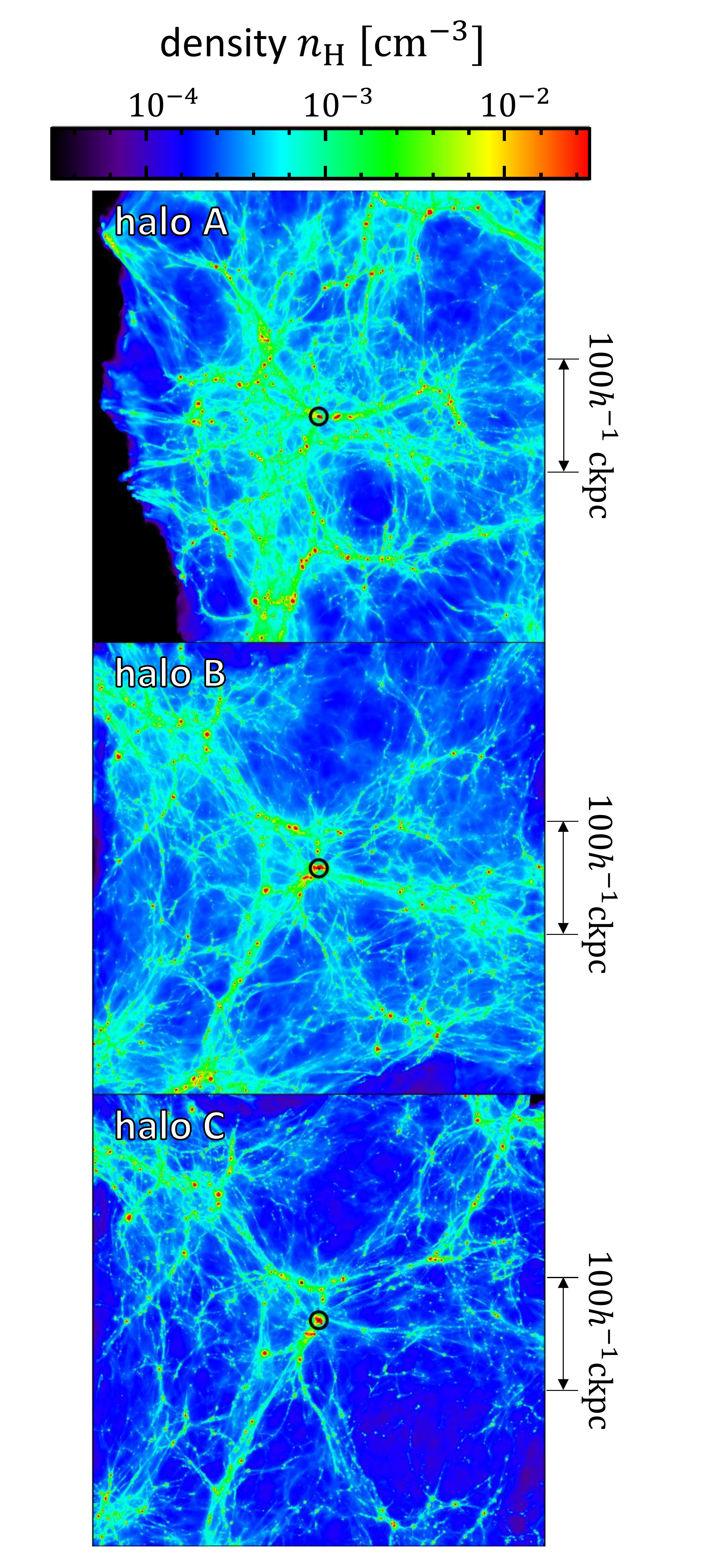}
\caption{Projected density maps in the zoom-in regions at the epoch of $z \sim 10$, for the cases of halos A, B, and C in descending order. In each panel, the central black circle denotes the virial radius of our target halo. 
    }
    \label{fig:splash_cosmological}
\end{figure}

Fig.~\ref{fig:splash_cosmological} displays the two-dimensional projected density maps in our zoom-in regions at the epoch of $z \sim 10$.
Our target halos A, B, and C are all located at the intersection of multiple filamentary structures, through which the halos accrete the DM and baryons. These most massive halos tend to reside in the densest part of each cosmological volume.


In Section~\ref{subsec:caseA} below, we first describe the simulation results, focusing on the case of halo A, where the halo grows earliest among the cases considered.
Since the density within the halo is proportional to the cosmological mean value $\bar{n}_{\rm H}\propto (1+z)^3$, it is most likely that the accretion flow causes dense shocks favoured for the SMS formation. In Section~\ref{subsubsec:pen1}, we present the cosmological evolution of the accretion flow toward halo A observed in our simulation. We consider the evolution of the representative shock position in Section~\ref{subsubsec:pen2}. In Section \ref{subsec:3cases}, we apply the same analyses for the other cases of halos B and C, showing that the evolution is qualitatively similar among all the cases. We further develop a semi-analytical model to interpret the results in Section \ref{subsec:interpretation}.


\subsection{Fiducial case of halo A}
\label{subsec:caseA}
\subsubsection{Emergence of cold accretion}
\label{subsubsec:pen1}

\begin{figure}
	\includegraphics[bb=0 0 430 320,width=10cm,scale=0.2]{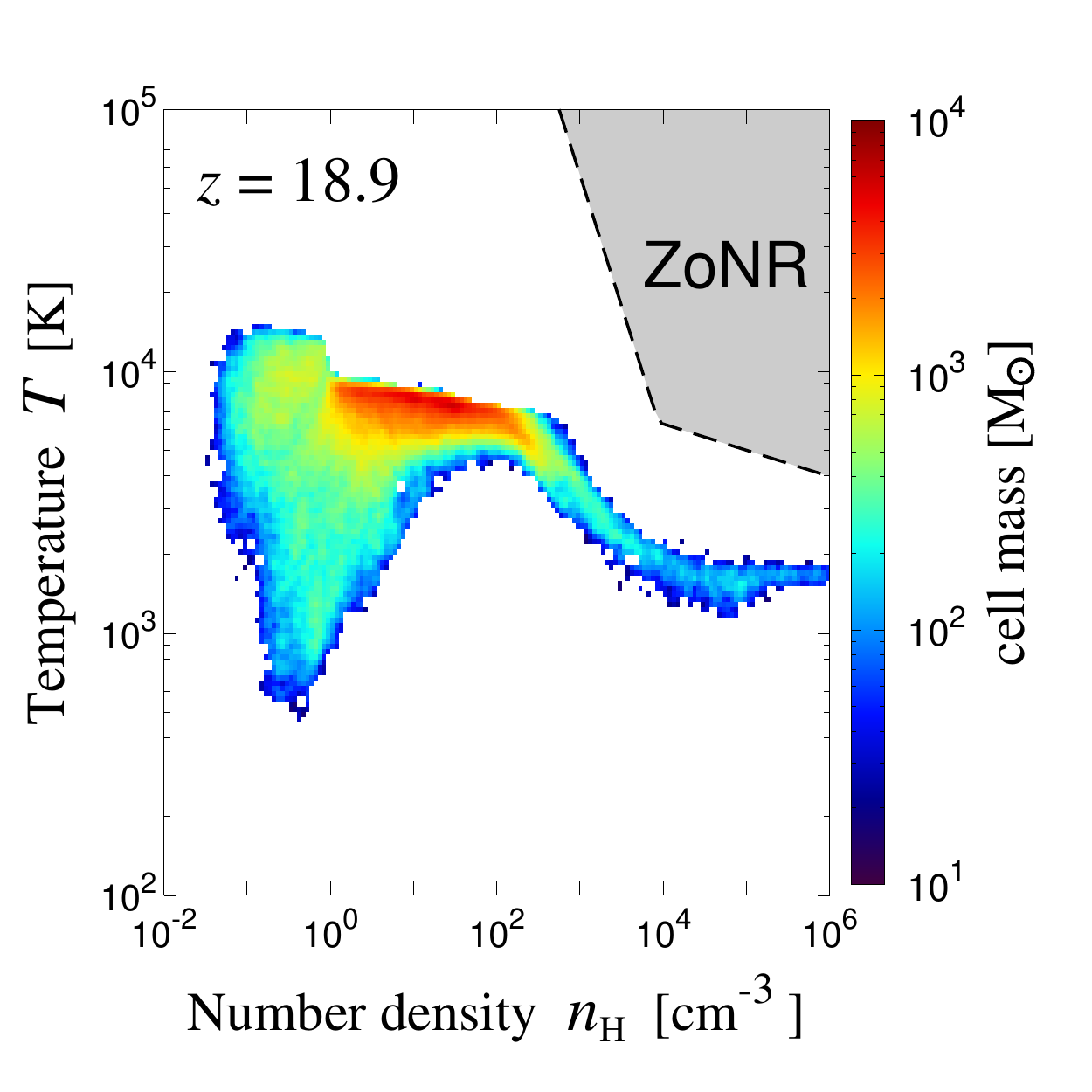}
    \caption{The mass distribution on the density-temperature plane within the virial radius of halo A at the epoch of $z=18.9$. The colour bar represents the particle mass in each bin. The gray area at the upper-right corner corresponds to the ZoNR given by \citet{InayoshiOmukai2012}.
    }
    \label{fig:rho_T_Fer}
\end{figure}

Because of the LW background field of $J_{21}=10$ we assume, the gravitational collapse of a cloud occurs after $T_\mathrm{vir}$ exceeds 10$^4$~K, relying on Ly$\alpha$ cooling in halo A. Fig.~\ref{fig:rho_T_Fer} shows the snapshot of the gas distribution on the density-temperature plane when the cloud's central density reaches $\sim 10^6~\cmc$ after the collapse. The gas is nearly isothermal at $T \simeq 8000\ {\rm K}$ for $n_{\rm H} \lesssim 10^2~\cmc$ owing to Ly$\alpha$ cooling. The temperature decreases down to $T\simeq 2000 \ {\rm K}$ for $n_{\rm H} \gtrsim 10^2~\cmc$, a signature that H$_2$ molecular cooling becomes effective. 
\citet{Fernandez2014} also show similar results.
The gas temperature does not lower $\sim 200$~K for $n_{\rm H} \gtrsim 10^2~\cmc$, because of the omission of the self-shielding effect on ${\rm H_2}$ photodissociation by LW radiation. 
This treatment does not alter the large-scale dynamics of the cold accretion. We also discuss our treatment later in Section \ref{subsubsec:dense_shock}.
To study further development of the accretion flow toward the halo, we follow the long-term evolution afterward by continuing the simulation.  

\begin{figure}
	\includegraphics[bb=0 0 1180 920,width=10cm,scale=0.2]{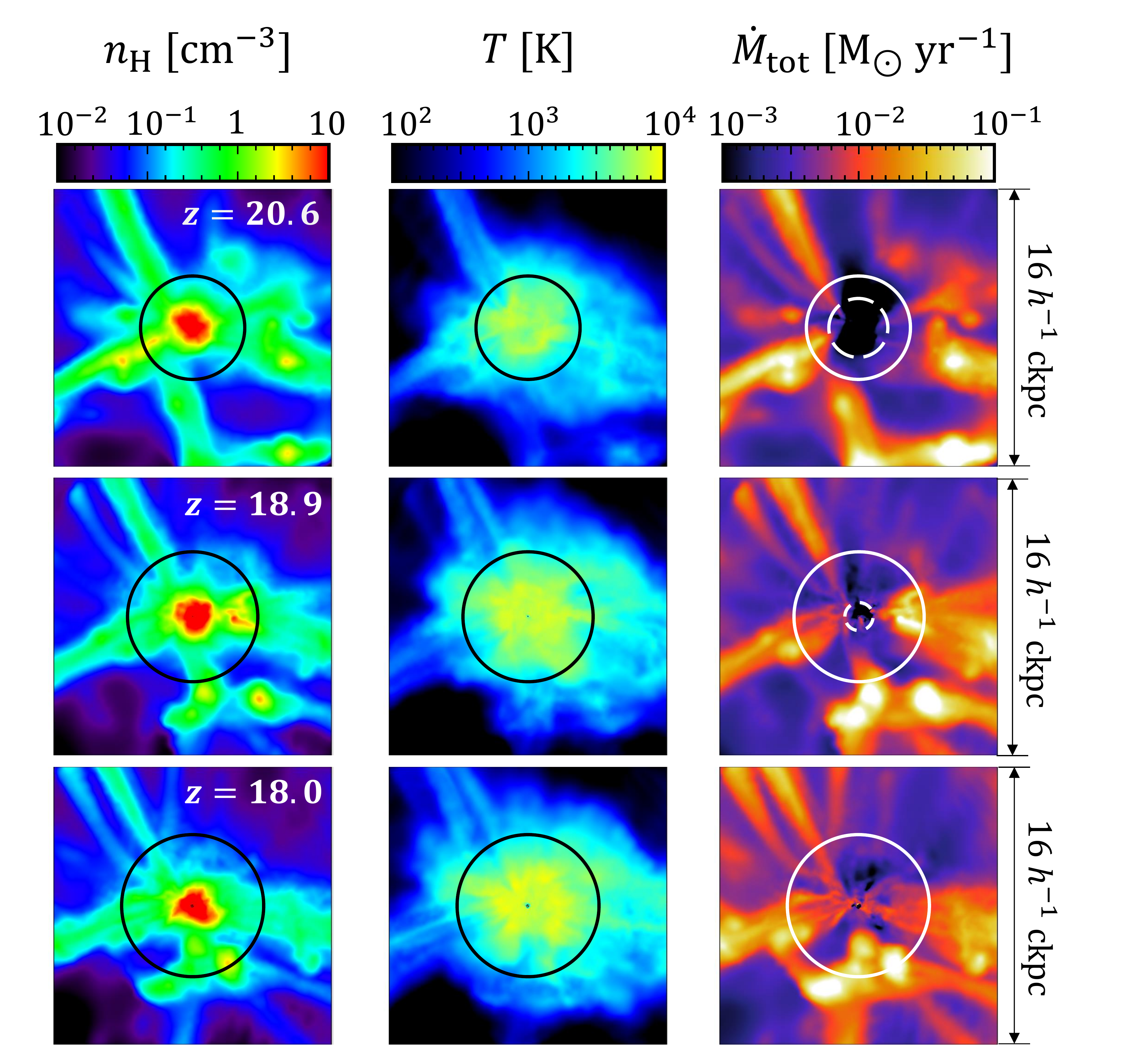}
    \caption{
Mass-weighted projection maps of halo A at different redshifts of $z=20.6$, 18.9, and 18.0 (top, middle, and bottom rows).
The epoch of the middle row corresponds to that of the snapshot in Fig.~\ref{fig:rho_T_Fer}. 
The left, middle, and right columns represent the gas density, temperature, and total mass accretion rate $\dot{M}_{\rm tot} \equiv \rho r^2v_{\rm inf}$ (also see Appendix~\ref{sec:shockposition}). The solid circle in each panel denotes the halo's virial radius. The dashed circles in the right column represent the shock radius $r_{\rm shock}$ evaluated by the method described in Appendix~\ref{sec:shockposition}.  
}
\label{fig:splash_Penetrate}
\end{figure}

Fig.~\ref{fig:splash_Penetrate} shows the temporal evolution of the accretion flow within $\sim 2 r_{\rm vir}$ in the case of halo A. The middle row presents the snapshots at the same epoch as in Fig.~\ref{fig:rho_T_Fer}. There is the large-scale filamentary accretion flow heading toward the halo centre. The bottom row presents snapshots after $\sim 10$~Myr, at the redshift $z = 18.0$. The right column shows the distribution 
of radial accretion rates\footnote{
Fig.~\ref{fig:splash_Penetrate} shows the total accretion rate $\dot{M}_{\rm tot}$ rather than that of only the fast component $\dot{M}_{\rm fast}$ (see also Appendix~\ref{sec:shockposition}). The total accretion rate is suitable for showing the accretion flow across the virial radius, particularly for the slow component outside the halo. }
$\dot{M}_{\rm tot}\equiv \rho r^2 v_{\rm inf}$, where $\rho$ is gas density, $r$ is radial distance from the halo center, and $v_{\rm inf}$ is radial infalling velocity.
Positive values of $v_{\rm inf}$ indicate the inward motion. 
This column indicates that the accretion flow penetrates the halo by the epoch of $z = 18.0$. The head of the accretion flow is at $r \simeq 0.1 r_{\rm vir}$ at $z=18.9$ (right middle panel), and it is further deep inside at $z = 18.0$ (right bottom panel).


The left and middle columns of Fig.~\ref{fig:splash_Penetrate} show the distributions of the density and temperature. In these panels, it is difficult to extract only the filamentary accretion flow within the virial radius. The gas temperature along the accretion columns is $\sim 10^4$~K, similar to the other virialized component. 
This is in stark contrast with the standard picture of the cold accretion \citep[e.g.][]{BirnboimDekel2003,Keres2005,Ocvirk+2008,Brooks+2009}, where the accretion flow with $T= 10^{4-5} \ {\rm K}$ penetrates into massive  ($M_\mathrm{halo} \sim 10^{10}~\Msun$) halos at low redshifts $z \simeq 0-4$. The accretion flow is much colder than the virialized gas component at $T_{\rm vir}=10^{6-7} \ {\rm K}$ for these cases. The high contrast in the temperature comes from the $T$-dependence of the cooling curve $\Lambda(T)$, which overall rises with decreasing $T$ in the range of $10^5~\mathrm{K} \lesssim T \lesssim 10^6~\mathrm{K}$. This leads to a significant decrease in temperature where cooling is efficient, such as in dense accretion filaments. For small halos with $M_{\rm halo}=10^{7-8}\ \Msun$, on the other hand, the virial temperature is $T_{\rm vir} \simeq 10^4\ {\rm K}$, where $\Lambda(T)$ takes peak values due to very efficient Ly$\alpha$ cooling. The middle column of Fig.~\ref{fig:splash_Penetrate} shows all the gas components within the virial radius have the same temperature of $T \simeq 8000 \ {\rm K}$, below which $\Lambda(T)$ drastically drops. In this perspective, it is misleading to use the term "cold accretion" for indicating the filamentary flow penetrating the halos with $M_{\rm halo} \sim 10^{7-8}\ \Msun$.  
We instead use "penetrating accretion" when necessary.  


\begin{figure}
	\includegraphics[bb=0 0 430 230,width=10cm,scale=0.2]{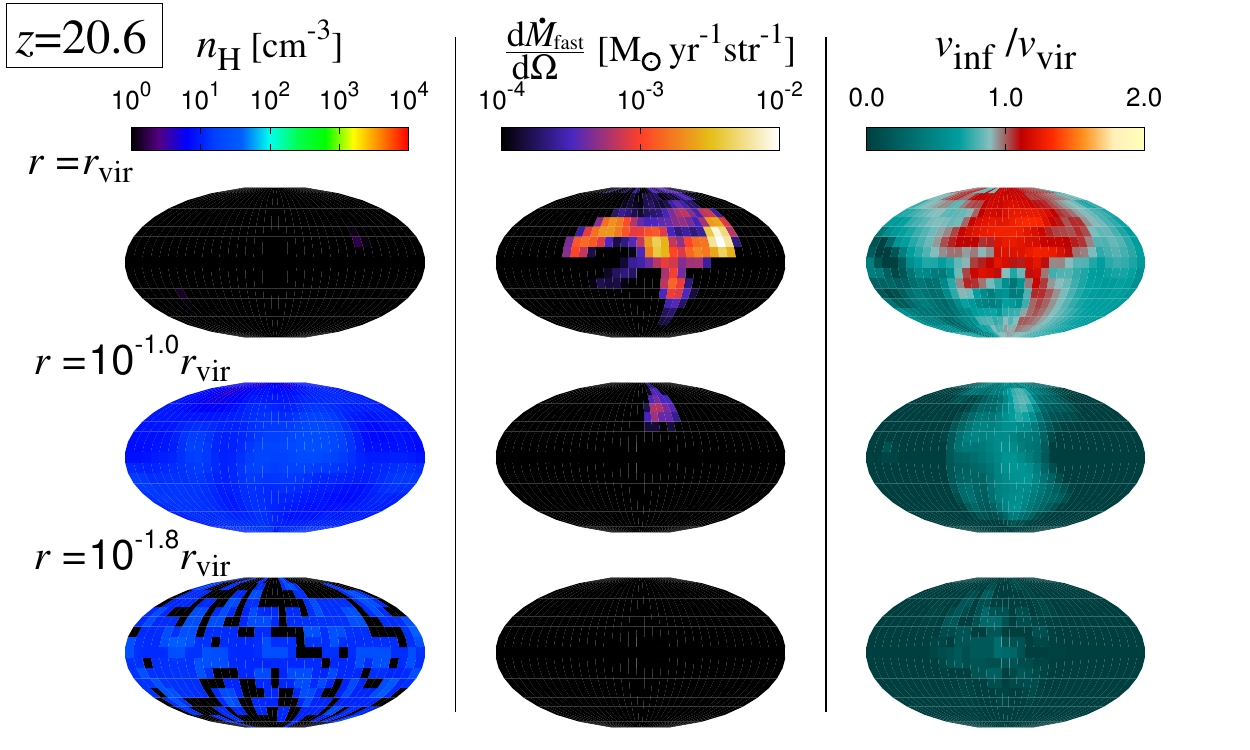}
\caption{
Mollweide projections of gas density (left column), gas mass accretion rate per unit solid angle (middle column),
    and infall velocity (right column) at $z = 20.6$ in halo A.
The top, middle, and bottom rows represent different radial slices,
$r=r_{\rm vir}$, $r=10^{-1.0}r_{\rm vir}$, and 
$r=10^{-1.8}r_{\rm vir}$, respectively.
}
\label{fig:Moll_089}
\end{figure}
\begin{figure}
	\includegraphics[bb=0 0 430 230,width=10cm,scale=0.2]{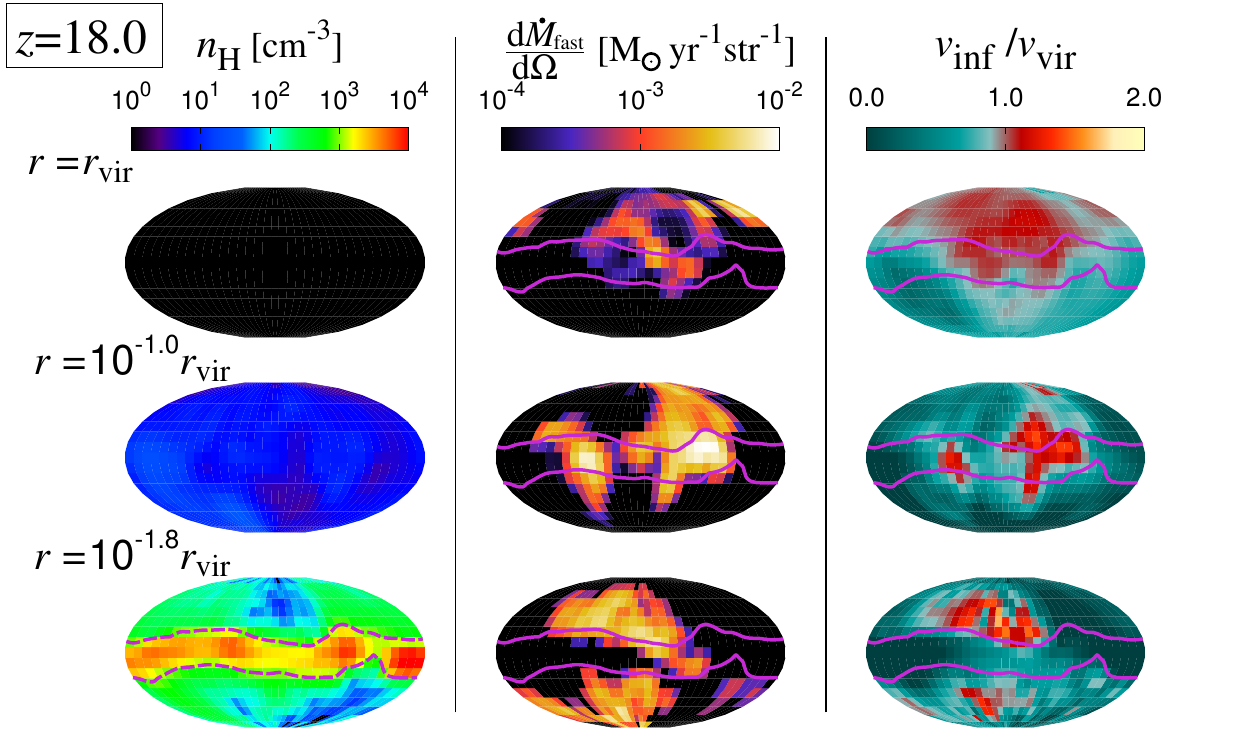}
    \caption{
    Same as Fig.~\ref{fig:Moll_089} 
    but for a later epoch of $z = 18.0$.
    The pink dashed lines in the bottom left panel represent an isodensity contour at $n_{\rm H}=10^3 \ \cmc$, corresponding to the central disc (see text). The pink solid lines in the middle and right columns are the same as that in the bottom left panel.
    }
\label{fig:Moll_017}
\end{figure}

Figs.~\ref{fig:Moll_089} and \ref{fig:Moll_017} show the angular distribution of the gas within halo A at the different epochs of $z=20.6$ and $18.0$, before and after the emergence of the penetrating accretion. 
In each figure, the different panel rows show slices at the different radii of $r=r_{\rm vir}, 10^{-1.0}r_{\rm vir}$, and  $10^{-1.8}r_{\rm vir}$ in the descending order. The three columns represent different physical quantities of the gas density $n_{\rm H}$ (left), the mass accretion rate 
of only the fast component\footnote{
The fast component here is defined as the gas with its infalling velocity larger than the virial velocity $v_{\rm inf}>v_{\rm vir}$.
This component corresponds to the penetrating accretion flow, the free-falling gas without experiencing shocks (see also Section \ref{subsubsec:pen2}).
}
per unit solid angle ${\rm d}\dot{M}_\mathrm{fast}/{\rm d}\Omega$ (middle), and the infalling velocity $v_{\rm inf}/v_{\rm vir}$ (right), where virial velocity is defined as $v_{\rm vir} \equiv \sqrt{GM_{\rm halo}/r_{\rm vir}}$. Fig.~\ref{fig:Moll_089} suggests that the fast accretion flow stalls at  $r \gtrsim 10^{-1.0} r_{\rm vir}$ by creating shock fronts. We only see its signature in the top row showing the slice at $r=r_{\rm vir}$. In contrast, Fig.~\ref{fig:Moll_017} shows that the fast accretion flow continues to the smallest scale of $r = 10^{-1.8} r_{\rm vir}$. The bottom row suggests that there is a disc-like structure and the accretion flow eventually hits the disc surface from oblique directions with respect to the equatorial plane of the disc.
The middle row shows accretion flow comes in wider directions some of which correspond to the equatorial plane of the disc. The disc prevents the flow in these directions from penetrating deeper than its size of $\simeq 0.05~r_{\rm vir}$.

\begin{figure}
\includegraphics[bb=0 0 630 1460,width=8cm,scale=0.2]{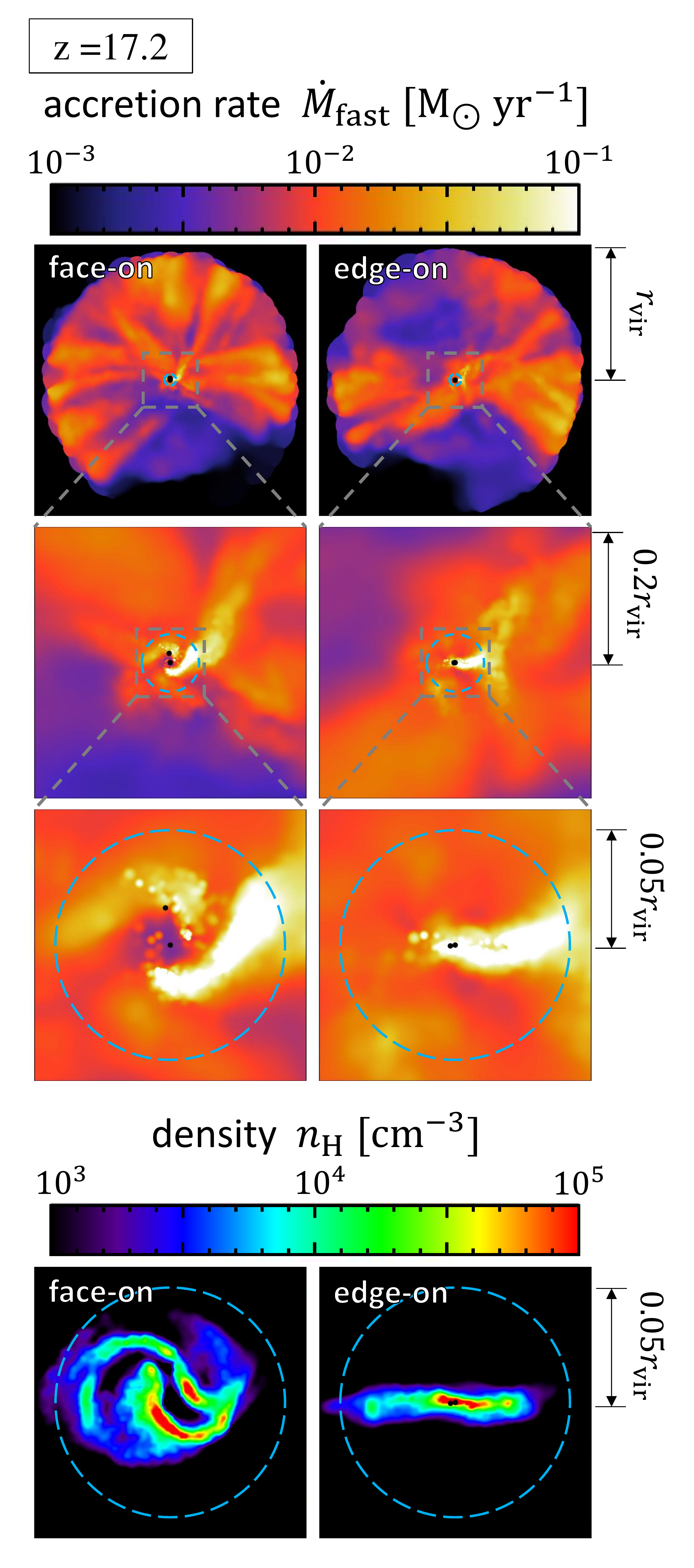}
\caption{
Accretion flow structure toward a central gas disc at the epoch of $z = 17.2$ in halo A. {\it Top three rows:} the density-averaged projection of the gas mass accretion rate $\dot{M}_{\rm fast}$, the contribution only from the fast component satisfying $v_{\rm inf} > v_{\rm vir}$.
The different rows provide close-up views of the small-scale structure, with the scales on the side of $10 h^{-1}\ {\rm ckpc}$, $2 h^{-1}\ {\rm ckpc}$, and $0.5 h^{-1}\ {\rm ckpc}$ in the descending order. The left and right columns correspond to the face-on and edge-on views with respect to the central gas disc. 
The light blue circles in the second and third rows represent the typical disc size, $r_{\rm disc}=0.05r_{\rm vir}$. The black dots represent sink particles. {\it Bottom row:} the gas density distribution within the central disc.
}
\label{fig:splash_Mdot}
\end{figure}

Fig.~\ref{fig:splash_Mdot} shows the structure of the gas accretion flow in a later epoch of $z=17.2$, at different spatial scales of $1~r_{\rm vir}$, $0.2~r_{\rm vir}$, and $0.05~r_{\rm vir}$. Note that we take the snapshots from an angle different from that in Fig.~ \ref{fig:splash_Penetrate}.
We see the structure of accretion flow and central disc, similar to that suggested by Fig.~\ref{fig:Moll_017}. There are multiple sink particles embedded within the disc.  
The face-on view (left column) is particularly informative to understand the flow structure. The main stream of the accretion flow comes from the upper right direction, which is evident at all the different spatial scales. This indicates that the filamentary accretion flow from the large-scale structure finally collides with the central disc.

\subsubsection{Time evolution of the shock position}
\label{subsubsec:pen2}

\begin{figure}
	\includegraphics[bb=0 0 950 500,width=11cm,scale=0.2]{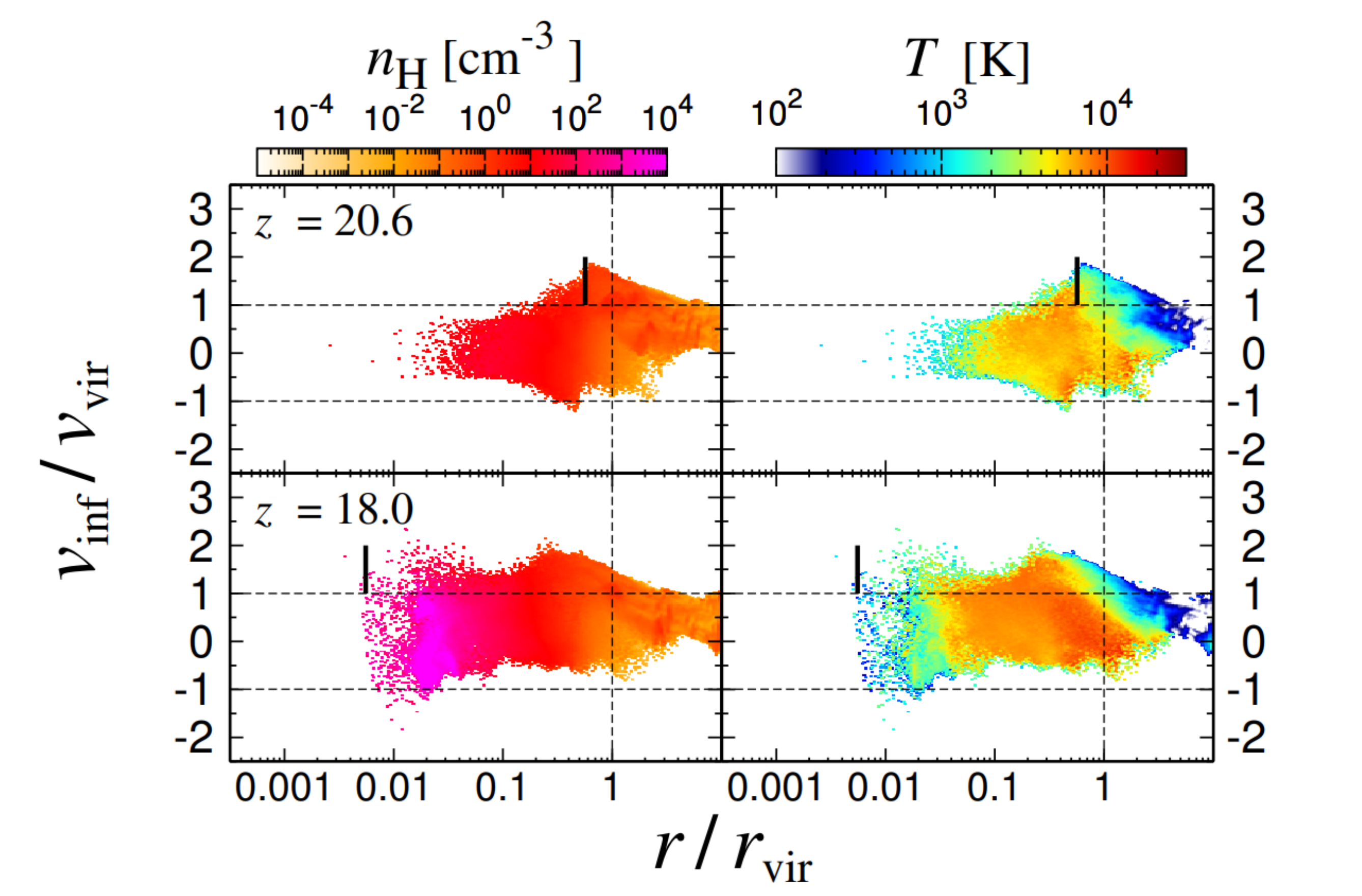}
    \caption{
Radial velocities and positions of the gas particles within $r < 10~r_\mathrm{vir}$ for the case of halo A. The horizontal and vertical axes represent the radial positions and velocities normalized by the virial values. Positive values of $v_\mathrm{inf}$ indicate the inward motion toward the halo centre. 
The upper and lower rows correspond to the different epochs of redshifts $z=20.6$ and $18.0$, respectively. The colours represent the gas number density (left column) and temperature (right column) of the particles.
The solid vertical bars represent the shock positions evaluated by the method described in Appendix~\ref{sec:shockposition}.
}
    \label{fig:r-v}
\end{figure}

We quantitatively consider the penetration of the accretion flow occurring in halo A. To this end, we evaluate the representative radius of the shock front $r_\mathrm{shock}$ for a given snapshot. We investigate the gas distribution in the phase space of the radial component, i.e., on the plane of the radial infall velocity $v_{\rm inf}$ against the radial position $r$ (see Appendix~\ref{sec:shockposition} for details).

Fig. \ref{fig:r-v} shows the scatter of the gas particles on the $r$-$v_{\rm inf}$ plane at the epochs of $z=20.6$ and 18.0 for the case of halo A. The gas in the area of $-1 \lesssim v_{\rm inf}/v_{\rm vir} \lesssim 1$ and $r \lesssim r_{\rm vir}$ corresponds to the virialized component.
There is also the additional component with the high infall velocity $v_{\rm inf}/v_{\rm vir} \gtrsim 1$ within the virial radius, which we consider the gas flowing into the halo being unshocked.
Such a component only distributes for $r \gtrsim 0.2~r_{\rm vir}$ at $z=20.6$, and it goes deeper inside for $r \gtrsim 0.005~r_{\rm vir}$ later at $z=18.0$. These features represent the penetration of the accretion flow. 

\begin{figure}
	\includegraphics[bb=0 0 330 300,width=9cm,scale=0.2]{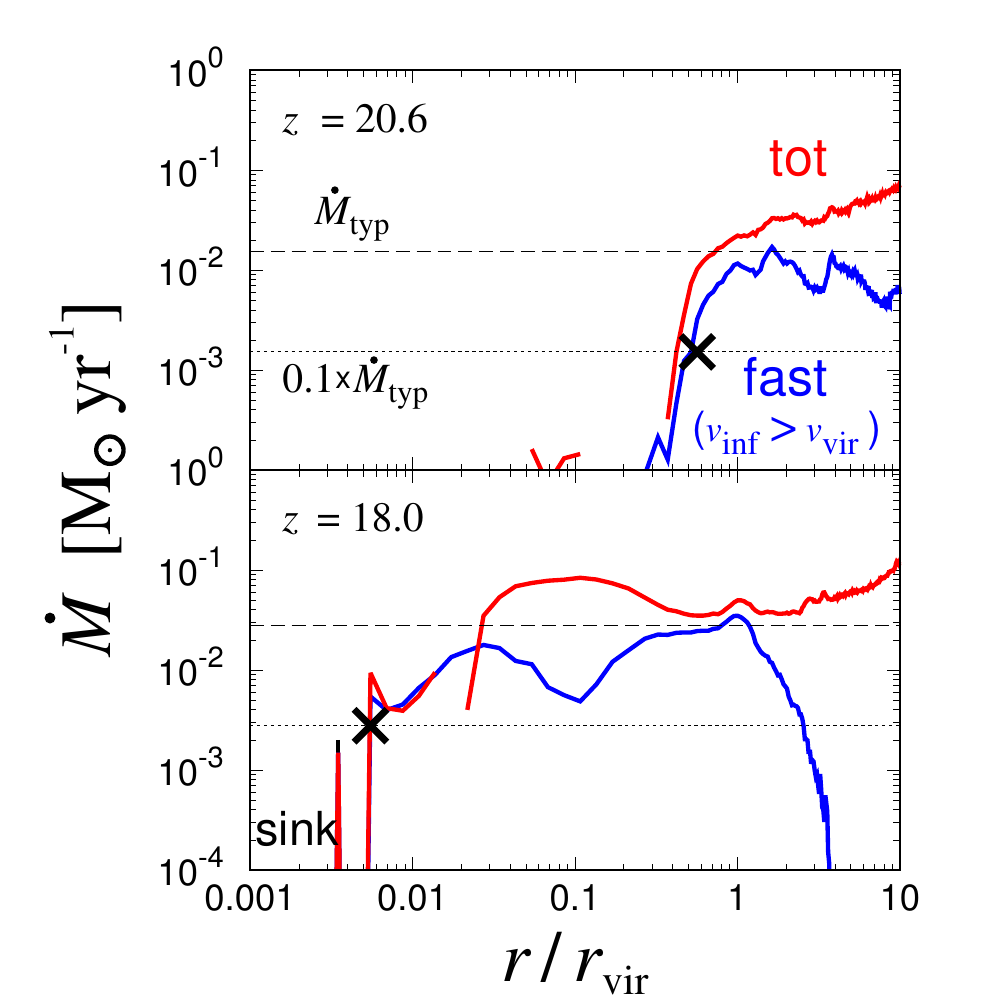}
    \caption{
Radial distribution of the gas mass accretion rates toward halo A at different epochs of the redshifts $z=20.6$ (upper panel) and $18.0$ (lower panel). The horizontal axis represents the radial distance from the halo centre normalized by the virial radius. In each panel, the red and blue lines represent the total accretion rates and those only with the fast component moving with the large infall velocity $v_{\rm inf}>v_{\rm vir}$. 
The horizontal dashed lines represent the typical rates $\dot{M}_{\rm typ}$ analytically estimated as functions of the halo mass $M_{\rm halo}$ and redshift $z$ (see Appendix \ref{sec:shockposition}). The cross symbols mark the positions where $\dot{M}$ for the fast component falls below $0.1 \dot{M}_{\rm typ}$, the representative shock positions $r_{\rm shock}$ we consider. 
}
    \label{fig:r-Mdot}
\end{figure}

Whereas Fig.~\ref{fig:r-v} indicates how the accretion flow develops within halo A, we quantitatively evaluate the representative shock position below. To this end, we calculate the mass accretion rates only by the fast component with $v_{\rm inf} > v_{\rm vir}$, $\dot{M}_{\rm fast}$, as a function of the radius $r$ (also see Appendix \ref{sec:shockposition}). Fig.~\ref{fig:r-Mdot} shows the results of such analyses, the radial distributions of $\dot{M}_{\rm fast}$ at the same epochs as in Fig.~\ref{fig:r-v} (blue lines). For instance, the upper panel of Fig.~\ref{fig:r-Mdot} shows that $\dot{M}_{\rm fast}$ sharply drops at $r \simeq 0.5r_{\rm vir}$ at the epoch of $z=20.6$, indicating that the fast accreting gas typically experiences the shock at that point. The lower panel shows that, at the later epoch of $z=18.0$, $\dot{M}_{\rm fast}$ takes $\sim 0.01~\mdotsunyr$ for $r \gtrsim 0.005 r_{\rm vir}$, corresponding to the size of the sink particle at the halo centre. These panels both show that $\dot{M}_{\rm fast}$ substantially decreases at a given radius by orders of magnitudes, allowing us to define the shock radius $r_{\rm shock}$ as follows. We consider the analytic formula of the cosmological mean accretion rate onto a halo $\dot{M}_{\rm typ}$, which well approximates the total gas accretion rates irrespective of the infall velocity $\dot{M}_{\rm tot}$ (red lines). We define $r_{\rm shock}$ as the innermost radius where $\dot{M}_{\rm fast} =0.1 \times \dot{M}_{\rm typ}$. We choose the factor of $0.1$ to capture the sharp drop of $\dot{M}_{\rm fast}$. Taking the smaller values does not change our results. 


The shock radius evaluated by the above method agrees with characteristic features in the simulation run. In Fig.~\ref{fig:r-v}, for instance, the vertical bars representing $r_{\rm shock}$ provide the lower bounds on the radial distribution of the fast-component particles with $v_{\rm inf}/v_{\rm vir} \gtrsim 1$. In Fig.~\ref{fig:splash_Penetrate}, moreover, the white dashed circle representing $r_{\rm shock}$ traces the head positions of the filamentary accretion columns within the virial radius.

\begin{figure*}
	\includegraphics[bb=0 0 780 270,width=20cm,scale=0.2]{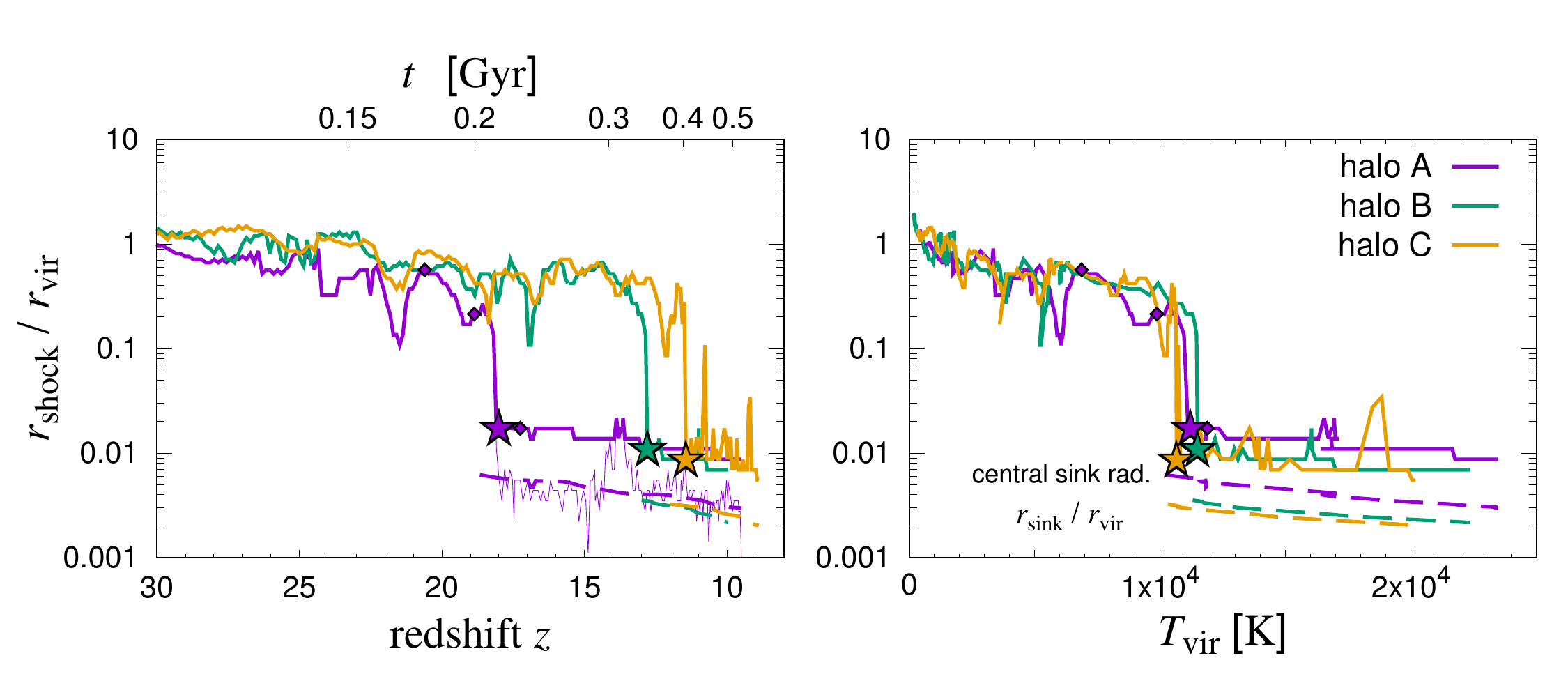}
\caption{
Cosmological evolution of the representative shock radius $r_\mathrm{shock}$ within halos A, B, and C (purple, green, and orange lines).  
The left and right panels show the evolution of $r_{\rm shock}$ normalized by the halo virial radius $r_{\rm vir}$ against the redshift (and cosmic age) and virial temperature $T_{\rm vir}\propto M_{\rm halo}^{2/3}(1+z)$.
The star symbols on the lines mark the epochs when $r_{\rm shock}$ drops below the typical disc size $r_{\rm disc} = 0.05~r_{\rm vir}$ for the first time. In the left panel, the purple diamonds denote the epochs of the snapshots shown in Fig.~\ref{fig:splash_Penetrate}.
In the left panel, the dashed lines for $z \lesssim 18$ represent the sink radius $r_{\rm sink}$ measured from the centre of halos, and the thin purple line shows our original estimates of $r_{\rm shock}$. Note that we impose the lower limit of $r_{\rm shock}$ as $3 r_{\rm sink}$ to avoid artificial effects of the sink (also see text). In the right panel, the dashed lines for $T_{\rm vir} \gtrsim 10^4$~K also represent the sink radii measured from the halo centres. 
}
\label{fig:z-Rshock}
\end{figure*}

Applying the above analysis to all the snapshots, we obtain the time evolution of the shock position $r_{\rm shock}$, as shown in Fig.~\ref{fig:z-Rshock}. In the left panel, our original estimates of $r_{\rm shock}$ provide the radius comparable to or even smaller than the sink radius $r_{\rm sink}$ for $z \lesssim 18$ (see purple dashed and thin solid lines). This is due to our prescription of sink particles. Since we only remove gas particles within the sink radius every few dozen timesteps, the shock radius can become smaller than the sink radius. We regard such very small $r_{\rm shock}$ as artifacts and impose the lower limits of $r_{\rm shock} \geq 3 r_{\rm sink}$. We re-define $r_{\rm shock}$ if it is less than $3 r_{\rm sink}$. These modified estimates of $r_{\rm shock}$ correspond to the thick solid line in Fig.~\ref{fig:z-Rshock}.  
Regardless of such technical details, the left panel of Fig.~\ref{fig:z-Rshock} shows a clear overall trend. The shock radius $r_{\rm shock}$ abruptly decreases at the redshift $z \simeq 18$, $\sim 10 \ {\rm Myr}$ after the first run-away collapse in the ACH,
when the halo mass is $M_{\rm halo} = 1.58 \times 10^7~\Msun$. This is the signature of the penetration of the accretion flow through halo A. 
Before the epoch of the penetration, the shock resides around the virial radius. There are only short periods when $r_{\rm shock}$ temporarily decreases to $\simeq 0.1 r_{\rm vir}$ at $z \gtrsim 18$. These correspond to halo major mergers, after which $r_{\rm shock}$ recovers to $\simeq 0.5 r_{\rm vir}$. After the epoch of $z \simeq 18$, the shock always stands at $r_{\rm shock} < 0.05r_{\rm vir}$, indicating that the penetrating accretion flow continues to hit the central gas disc until the end of the simulation at $z \simeq 10$.

\subsection{General trends in cases of halo A, B, and C}
\label{subsec:3cases}

In addition to the fiducial case of halo A, we apply the same analysis as in Section~\ref{subsubsec:pen2} to the other cases of halos B and C. Although not presented, we obtained similar evolution as in Figs.~\ref{fig:r-v} and \ref{fig:r-Mdot} for these cases. Fig.~\ref{fig:z-Rshock} also shows the resultant cosmological evolution of $r_{\rm shock}$ for the cases of halos B and C. 
The time evolution of $r_{\rm shock}$ in all the cases share common features, the sharp drop from $r_{\rm shock} \simeq 0.5 r_{\rm vir}$ to $\sim 0.01 r_{\rm vir}$. That is, the penetration of the accretion flow generally occurs for relatively short periods, at some point during $10 < z < 20$.
After such critical redshifts, the fast accretion flow reaches $\sim 0.01 r_{\rm vir}$, directly hitting a gas disc deeply embedded at the centre of each halo. 
In the cases of halo A, B, and C, the penetration occurs $\sim 10-30~{\rm Myr}$ after the first run-away collapse in ACHs.


To understand what determines the epoch when the accretion flow penetrates the halo, we consider the evolution of $r_\mathrm{shock}$ against the virial temperature $T_{\rm vir}\propto M_{\rm halo}^{2/3} (1+z)$. Since $M_{\rm halo}$ rapidly increases by a factor of $\sim 100$ from $z= 20$ to $10$, $T_{\rm vir}$ is an increasing function of the cosmic time, overwhelming the dependence of $T_{\rm vir} \propto (1+z)$.
The right panel of Fig. \ref{fig:z-Rshock} shows that the epochs of the sharp drop of $r_{\rm shock}/r_{\rm vir}$ are almost identical at $T_{\rm vir} = 1.1 \times 10^4\ {\rm K}$ for all the cases of halo A, B, and C. 
This is reasonable because radiative cooling is most efficient owing to strong Ly$\alpha$ emission around $\sim 10^4$~K. Below in Section~\ref{subsec:interpretation}, we also provide semi-analytic modeling for interpreting such critical behaviour of the accretion flow across $T_{\rm vir} \simeq 10^4\ {\rm K}$.

\begin{figure}
	\includegraphics[bb=0 0 530 350,width=10cm,scale=0.2]{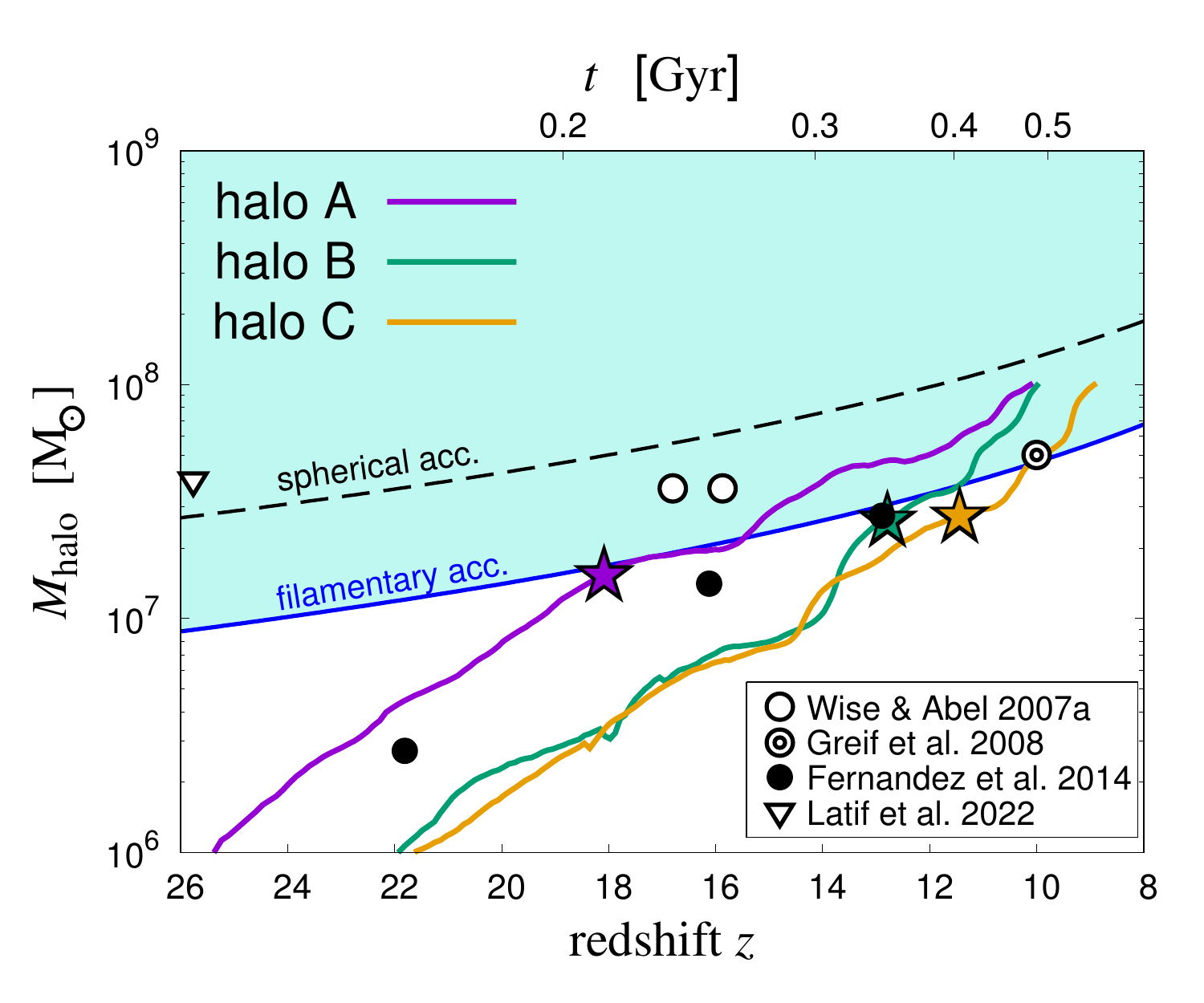}
    \caption{
The first emergence of the penetrating (or cold) accretion flow in the halo mass assembly histories. The horizontal axis represents the redshift (and cosmic age), and the vertical axis represents the halo mass $M_{\rm halo}$. The purple, green, and orange lines represent the cases of halo A, B, and C, respectively. The star symbols on the lines indicate the same epochs as in Fig.~\ref{fig:z-Rshock}, when the accretion flow reaches the hearts of the halos. The blue solid and black dashed lines represent our semi-analytical evaluation of the critical halo masses above which the penetrating accretion flow is possible. The spherically symmetric accretion and filamentary accretion are assumed for deriving the black and blue critical lines (see Section~\ref{subsec:interpretation}). 
The open, double, and filled circles and 
 inverted triangle
represent the results of previous simulations reported by \citet{WiseAbel2007}, \citet{Greif2008}, \citet{Fernandez2014}, and \citet{Latif2022}.
The filled circles represent cases where cold accretion did not occur, and open circles represent cases where accretion reached about half the virial radius. The double circles represent cases of complete penetration.
}
\label{fig:massevo}
\end{figure}

Fig. \ref{fig:massevo} summarizes when the accretion flow reaches the halo centres in the halo assembly histories. The halo masses for the first emergence of the penetrating accretion are written as 
\begin{eqnarray}
M_{\rm halo,min}&\simeq& 2.20\times 10^7 \ {\rm M}_\odot \left(\frac{1+z}{15}\right)^{-3/2},
\label{eq:mhalocr}
\end{eqnarray}
which corresponds to the virial tempearture of $T_{\rm vir}=1.1\times 10^4\ {\rm K}$. The accretion flow continues to hit the central disc after the halo mass exceeds the above value. In Fig.~\ref{fig:massevo}, the circle symbols represent the final snapshots of previous relevant simulations studying the cold accretion in small halos. 
For instance, the filled circles represent the three cases considered in \citet{Fernandez2014}, who report that the accretion flow hardly enters deep inside the virial radius. 
Fig. \ref{fig:massevo} shows that our derived minimum halo masses are heavier than their final ones. The double circle represents the case of \citet{Greif2008}, who demonstrate that the accretion flow reaches the halo centre in their simulation. The halo mass at their final snapshot is very close to the value of Equation~\eqref{eq:mhalocr}. 
\citet{Latif2022}, denoted by the inverted triangle, report that the gas accretion flows have reached
$\sim 10 \ {\rm pc} \ \sim 0.01 r_{\rm vir}$ from the halo center by $z=25$.
They mainly analyse two epochs of around $z=29,25$.
As we cannot tell whether the flows penetrate or not at other epochs, we show the symbol as an upper limit of the minimum halo mass with which the flows penetrate.
\citet{WiseAbel2007}, denoted by the open circles, report that the gas accretion flow reaches the depths of about half of the virial radius.
In Section~\ref{subsec:diff}, we further discuss our results in comparison to these previous studies considering differences in the simulation and analysis methods in more detail.

\subsection{Interpreting simulations with semi-analytical modeling}
\label{subsec:interpretation}

\citet{BirnboimDekel2003} developed a semi-analytic model of the cold accretion assuming spherical symmetry. The model provides the condition for the accretion flow to penetrate a halo centre. The model predicts that the cold accretion should appear {\it below} the critical halo mass $M_{\rm halo}=10^{10-12}\ \Msun$ supposing massive galaxy formation at low redshifts. In contrast, our simulations suggest that the accretion flow begins to reach the halo centres {\it above} the halo mass given by Equation~\eqref{eq:mhalocr}. In this section, we apply the semi-analytic modeling to interpret our simulation results.


The model by \citet{BirnboimDekel2003} is briefly outlined as follows. Suppose that a thin gas spherical shell free-falls onto a halo with the mass $M$ and experiences a shock at the radius $r=r_{\rm vir}$ at the redshift $z=z_{\rm vir}$.
The density within the shell is estimated by considering the radial motion of the two spherical shells with the enclosed mass $M$ and $M+{\rm d}M$ as 
\begin{eqnarray}
\rho\equiv \frac{f_{\rm br}{\rm d}M}
{4\pi r(M)^2\left\{r(M+{\rm d}M)-r(M)\right\}},
\end{eqnarray}
where $f_{\rm br}=\Omega_{\rm br}/(\Omega_{\rm DM}+\Omega_{\rm br})$ is the baryon fraction. We evaluate whether the post-shock spherical shells remain at $r_{\rm vir}$ or continue to free-fall in the following way. We take the infall velocity, gas density, and temperature just before the shock as the pre-shock values $u_{\rm pre}$, $\rho_{\rm pre}$, and $T_{\rm pre}$, and apply the adiabatic Rankine-Hugoniot jump conditions to obtain the post-shock values $u_{\rm post}$, $\rho_{\rm post}$, and $T_{\rm post}$ as
\begin{eqnarray}
  u_{\rm post}&=&
  \left[\frac{\gamma-1}{\gamma+1}+\frac{2}{\gamma+1}{\mathscr M}^{-2}\right]
  u_{\rm pre},  \label{eq:RU_u}\\
  \rho_{\rm post}&=&
  \left[\frac{\gamma-1}{\gamma+1}+\frac{2}{\gamma+1}{\mathscr M}^{-2}\right]^{-1}
  \rho_{\rm pre},  \label{eq:RU_n}\\
  T_{\rm post} &=& \frac{\left\{2\gamma-(\gamma-1){\mathscr M}^{-2}\right\}
  \left\{(\gamma-1)+2{\mathscr M}^{-2}\right\}}{(\gamma+1)^2}
   {\mathscr M}^2T_{\rm pre}, \nonumber\\
   &&\label{eq:RU_T}
\end{eqnarray}
where ${\mathscr M}$ is the pre-shock Mach number
\begin{eqnarray}
  {\mathscr M}^{2} &\equiv& \frac{u_{\rm pre}^2}{c_{\rm s,pre}^2}
  =\frac{\mu m_{\rm H}u_{\rm pre}^2}{\gamma k_{\rm B}T_{\rm pre}},\label{eq:RU_M}
\end{eqnarray}
$\mu$ the mean molecular weight, and $\gamma$ the adiabatic index. Assuming that the gas temperature in the pre-shock state is much lower than the post-shock value and that the strong-shock approximation $\mathscr{M}\gg 1$ is fulfilled within the zeroth order of $\mathscr{M}^{-2}$, Equations \eqref{eq:RU_u} - \eqref{eq:RU_T} become 
\begin{eqnarray}
  u_{\rm post}&\simeq&
  \frac{\gamma-1}{\gamma+1}u_{\rm pre}
  =\frac{1}{4}u_{\rm pre}, \label{eq:RU_u2}\\
  \rho_{\rm post}&\simeq&
  \frac{\gamma+1}{\gamma-1}
  \rho_{\rm pre} =4\rho_{\rm pre}, \label{eq:RU_n2}\\
  T_{\rm post} &\simeq& \frac{2\gamma(\gamma-1)}{(\gamma+1)^2}
   \frac{\mu m_{\rm H} u_{\rm pre}^2}{\gamma k_{\rm B}}
   =\frac{3}{16}\frac{\mu m_{\rm H} u_{\rm pre}^2}{k_{\rm B}}, \label{eq:RU_T2}
\end{eqnarray}
where we use $\gamma=5/3$. We then compare the following timescales: the kinetic timescale $t_{\rm dyn}\equiv r_{\rm vir}/u_{\rm post}$ and the cooling timescale
\begin{eqnarray}
t_{\rm cool}&\equiv&\frac{\frac{3}{2}n_{\rm H,post}k_{\rm B}T_{\rm post}}{\rho_{\rm post}^2\Lambda(T_{\rm post})},
\end{eqnarray}
where $\Lambda (T)$ is the cooling function at zero metallicity.  If $t_{\rm dyn}\ll t_{\rm cool}$, the thermal pressure balances with the gravity in the post-shock layer, resulting in the gas shell staying at $r_{\rm vir}$. If $t_{\rm dyn}\gg t_{\rm cool}$, the thermal energy at the post-shock region is reduced rapidly by radiative cooling. In this case, the shock front no longer stays around the virial radius, and the accretion flow gets deeper into a halo at the supersonic velocity. 
We regard $t_{\rm dyn}=t_{\rm cool}$ as the condition dividing whether the gas flow stalls at $r_{\rm vir}$ or penetrates toward the halo centre. We apply the above model to small halos we consider, adopting the same fiducial model parameters as in \citet{BirnboimDekel2003}.


The black dashed line in Fig.~\ref{fig:massevo} represents the minimum halo masses given by the semi-analytic model, above which the accretion flow plunges deep into the halo. The minimum halo mass is approximated as $M_{\rm crit} \simeq 7.6\times 10^7\ {\rm M}_\odot \left\{(1+z)/15 \right\}^{-3/2}$, for which the corresponding virial temperature is $T_{\rm vir}\simeq 2.6\times 10^4\ {\rm K}$.
Fig.~\ref{fig:massevo} shows that the critical halo masses provided by \citet{BirnboimDekel2003} model roughly match our simulation results of minimum halo mass represented with star symbols, albeit about $3$ times larger than our simulation results. We further mitigate the discrepancy as follows.


Fig.~\ref{fig:splash_Penetrate} suggests that the geometry of the accretion flow is far from the spherical symmetry. Most of the infalling gas comes into a halo through the filamentary cosmic web, 
where the density and temperature are higher than the surrounding medium by a few orders of magnitude.
For a given pre-shock velocity $u_{\rm pre}$,
increasing the pre-shock density raises the post-shock density, resulting in shortening the cooling time through the dependency of $t_{\rm cool}\propto \rho_{\rm post}^{-1}$.
Increasing the pre-shock temperature raises the post-shock temperature by a small factor, resulting in
substantially enhancing cooling rate $\Lambda(T_{\rm post})$ in a post-shock layer due to its strong $T$-dependence around $T_{\rm post}\simeq 8000 \ {\rm K}$.
These result in shortening cooling timescale $t_{\rm cool}$, or facilitating the penetration of the accretion flow. We confirm that in our simulations the density and temperature in the filamentary accretion flow measured at $r=r_{\rm vir}$ are much higher than those supposed in the spherical model by \citet{BirnboimDekel2003}. To consider such filamentary accretion, we modify the model in the following manner. We take the density and temperature within the filamentary flow in our simulations and use them as the pre-shock quantities $n_{\rm pre}$ and $T_{\rm pre}$ in the model.  
To do so, we choose the snapshots at the epochs marked by the star symbols in Fig.~\ref{fig:massevo}. 
Table~\ref{table:filament} summarizes the mass-weighted mean values of the density and temperature of the fast component at $r=r_{\rm vir}$, $n_{\rm H,sim}$ and $T_{\rm sim}$, and the density in the spherical model as references.
We fit the three data sets of $(z,n_{\rm H,sim})$ by the function of $n_{\rm H,sim}={\rm const.}\times (1+z)^3$ and obtain $n_{\rm H, pre}=0.32 \ {\rm cm^{-3}} \{(1+z)/13\}^3$, for which the standard deviation is $0.17 \ {\rm cm^{-3}}$. We simply average the three values of $T_{\rm sim}$ and get $T_{\rm pre}=2510\ {\rm K}$, for which the standard deviation is $483 \ {\rm K}$.

\begin{table}
  \caption{Physical properties of the accretion flow at the virial radius given by the semi-analytic model and actual simulations.}
  \label{table:filament}
  \centering
  \begin{tabular}{l|cccc}
    \hline
    & $z$  &  $n_{\rm H, model} \ [{\rm cm^{-3}}]$ &$n_{\rm H,sim} \ [{\rm cm^{-3}}]$ &$T_{\rm sim} \ [{\rm K}]$\\
  \hline \hline
  A& $18.0$  & $3.64\times 10^{-2}$ & $1.074$ & $1.89\times 10^3$ \\
  B& $12.8$  & $1.21\times 10^{-2}$ & $0.120$ & $3.07\times 10^3$ \\
  C& $11.4$  & $8.81\times 10^{-3}$ & $0.378$ & $2.56\times 10^3$ \\
    \hline
  \end{tabular}
\end{table}

Our simulations show that the actual Mach number in the filamentary accretion flow is $\mathscr{M}^{-2} \simeq 0.2$, for which the strong-shock approximation should be also improved. Using $u_{\rm pre}(r=r_{\rm vir})^2 = u_{\rm vir}^2=2 k_{\rm B}T_{\rm vir}/\mu m_{\rm H}$, we approximate Equations~\eqref{eq:RU_u} - \eqref{eq:RU_T} to the first order of $\mathscr{M}^{-2} = \gamma T_\mathrm{pre}/2 T_\mathrm{vir}$ as
\begin{eqnarray}
u_{\rm post}&=&\left[\frac{\gamma-1}{\gamma+1}+\frac{\gamma}{\gamma+1}\frac{T_{\rm pre}}{T_{\rm vir}}\right]u_{\rm pre},\label{eq:u_mod}\\
\rho_{\rm post}&\simeq&
\left[\frac{\gamma+1}{\gamma-1}-\frac{\gamma(\gamma+1)}{(\gamma-1)^2}
\frac{T_{\rm pre}}{T_{\rm vir}}\right]\rho_{\rm pre},\label{eq:rho_mod}\\
T_{\rm post}&\simeq& \frac{4(\gamma-1)}{(\gamma+1)^2}T_{\rm vir}
+\frac{4\gamma-(\gamma-1)^2}{(\gamma+1)^2}T_{\rm pre}, \label{eq:T_mod}
\end{eqnarray}
for which we use $\gamma=5/3$.
Substituting $n_{\rm H, pre}$ and $T_{\rm pre}$ into these equations, we obtain the post-shock quantities to evaluate $t_{\rm dyn}$ and $t_{\rm cool}$. The equality of $t_{\rm dyn}= t_{\rm cool}$ gives the minimum halo masses. 


The solid blue line in Fig. \ref{fig:massevo} represents the result of our improved semi-analytic modeling.
In this case, the critical halo mass is approximated as $M_{\rm crit}\simeq2.9\times 10^7 \ {\rm M}_\odot \left\{ (1+z)/15 \right\}^{-3/2}$, corresponding to the virial temperature of $T_{\rm vir}\simeq 1.1\times 10^4\ {\rm K}$. The minimum halo masses considering the filamentary accretion flow are in very good agreement with those obtained from our simulations,
and also consistent with the results by \citet{Greif2008}, \citet{Fernandez2014}, 
and \citet{Latif2022}, 
despite our crude estimates of $n_{\rm H, pre}$ and $T_{\rm pre}$.
The cold accretion does not appear in \citet{Fernandez2014} because they terminate the simulation before it emerges. \citet{WiseAbel2007} show that the accretion flow reaches half of the virial radius in halos more massive than our minimum masses (see Section~\ref{subsec:diff} for further discussions).


\section{POSSIBILITY OF SUPERMASSIVE STAR FORMATION}
\label{sec:SMS_pre}

Following Section \ref{sec:results}, where we have studied the first emergence of the penetrating accretion flow in ACHs, we consider whether it leads to the SMS formation.
To this end, we study the supersonic accretion flow joining the central disc in more detail. We provide a maximal estimate of the dense and hot gas in the ZoNR created at dense shocks by such accretion flows. In Section \ref{subsec:haloA2}, we first look into the case of halo A. In Section \ref{subsec:3halos2}, we next apply the same analyses for the other cases of halos B and C, showing the general trends among all the cases.

\subsection{Fiducial case of halo A}
\label{subsec:haloA2}

\subsubsection{Dense shock by the penetrating accretion}
\label{subsubsec:dense_shock}

\begin{figure}
\includegraphics[bb=0 0 830 650,width=10cm,scale=0.2]{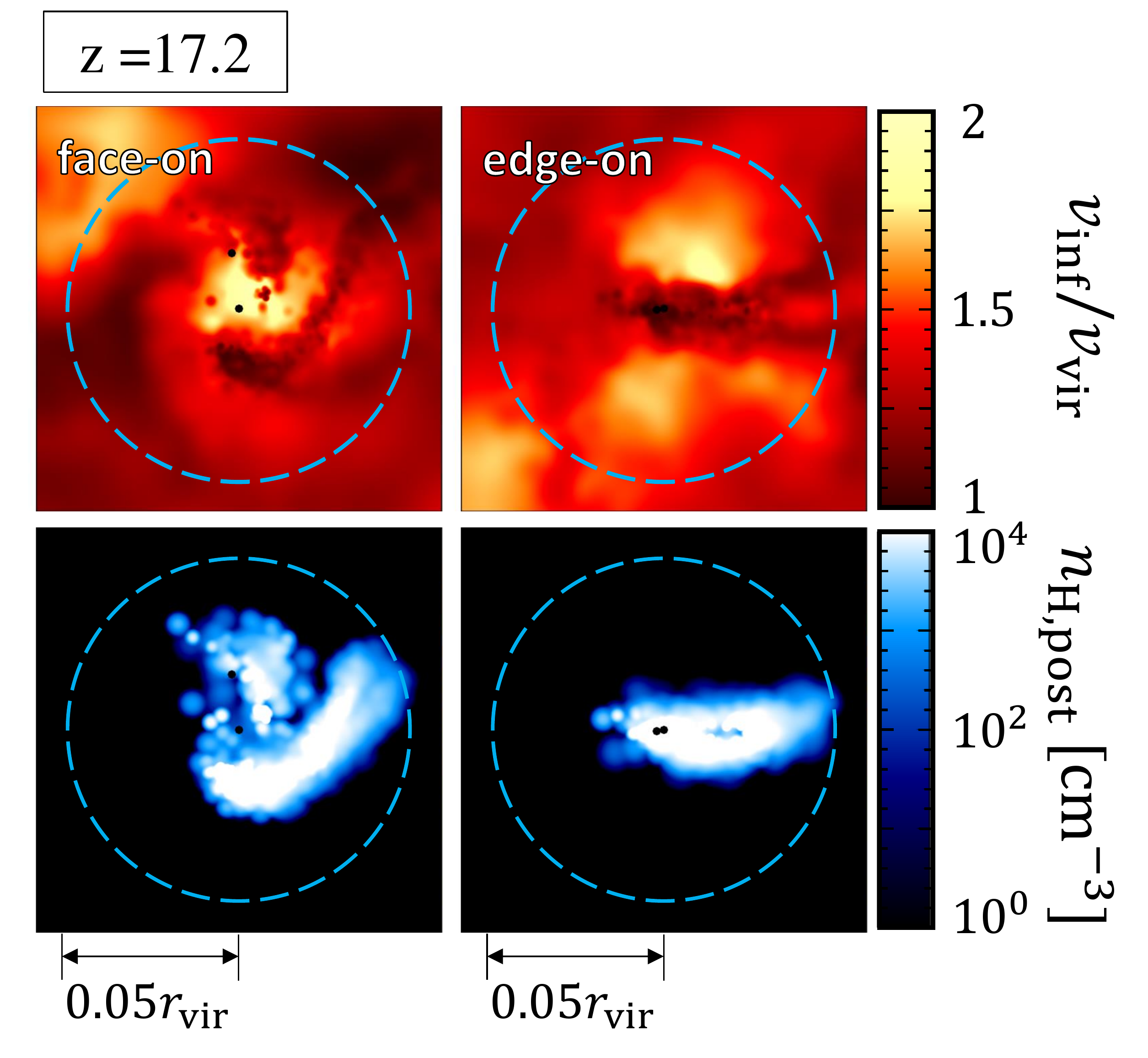}
    \caption{
Accretion flow toward the disc at the center of halo A and the spatial distribution of the dense hot gas. The left and right columns of the panels show the face-on and edge-on views with respect to the central disc at the epoch of $z=17.2$. The spatial scale and viewing angles are the same as in the bottom panels in Fig.~\ref{fig:splash_Mdot}. The upper and lower rows show the projected distributions of the infall velocities normalized by the virial value and post-shock density $n_{\rm H,post}$, which is calculated by shock jump conditions from the actual gas properties (see text). The lower panels only show the gas whose inferred post-shock state enters the ZoNR. The dashed circle in each panel represents the typical disc radius of $r=0.05 r_{\rm vir}$.
}
\label{fig:splash_others}
\end{figure}

We investigate the same snapshot at $z=17.2$ as in Fig.~\ref{fig:splash_Mdot} to study the accretion flow structure deep inside the halo A. The top panels of Fig. \ref{fig:splash_others} illustrate the projection maps of the infall velocity $v_{\rm inf}$ of the fast components. Fig.~\ref{fig:splash_others} shows two different accretion streams with high infall velocities. One comes from the polar directions toward the disc centre. This component has the relatively high velocity $v_{\rm inf} > 1.5 v_{\rm vir}$, but it only provides a minor contribution in terms of the accretion rate, $\dot{M}\sim 10^{-2} \ \mdotsunyr$. 
The other comes from equatorial directions along with the dense filaments with the slower velocities $v_{\rm inf}=(1.0-1.5) v_{\rm vir}$, and it has the higher mass accretion rate of $\dot{M}\sim 0.1 \ \mdotsunyr$. These streams originate from the same larger-scale filamentary accretion flow, which splits at some point before reaching the central disc. 

\begin{figure}
	\includegraphics[bb=0 0 400 350,width=10cm,scale=1.0]{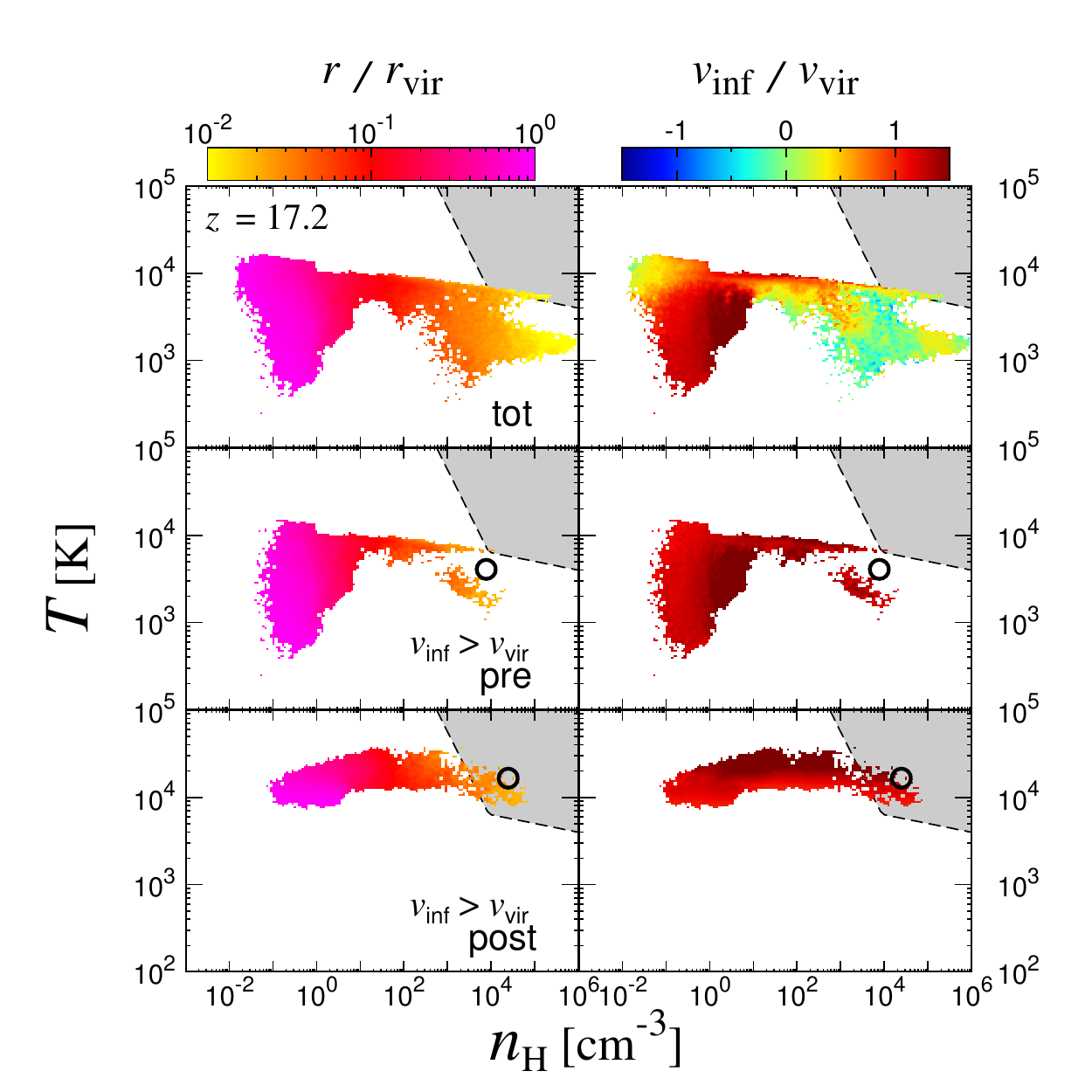}
    \caption{
Density-temperature distributions of the gas within the virial radius of halo A at the epoch of $z=17.2$. In the left and right columns of panels, different colours represent the distance from the halo centre and radial velocity normalized by the virial values, $r/r_\mathrm{vir}$ and $v_\mathrm{inf}/v_\mathrm{vir}$. Positive values of $v_\mathrm{inf}$ indicate the inward motion. The top and middle rows present the distributions of all particles and only those with high infall velocities $v_{\rm inf}>v_{\rm vir}$ (or fast component), respectively.
The bottom row shows the post-shock thermal states obtained by jump conditions assuming all the gas particles with $v_{\rm inf}>v_{\rm vir}$ instantly experience the shock in situ. In each panel, the gray area in the upper right corner represents the Zone of No Return, the same as in Fig.~\ref{fig:rho_T_Fer}.
The open circles in the middle and bottom rows represent the mass-weighted mean values of the fast component measured at representative shock radius $r_{\rm shock}$.
}
    \label{fig:rho_T}
\end{figure}

The top panels in Fig.~\ref{fig:rho_T} show the gas distribution within the virial radius on the density-temperature plane at the same epoch. The dense components with $n_\mathrm{H} \gtrsim 10^4~\cmc$ correspond to the central disc and its vicinity. Whereas most of the disc gas distributes below the ZoNR at $T \sim 10^3$~K, there is some amount of the gas within the ZoNR, near the lower boundary at $T \simeq 5000$~K. We regard this as the gas created by dense shocks at the central disc, as it is located at $r \lesssim 0.05 r_{\rm vir} \simeq r_{\rm disc}$. Its velocity $v_{\rm inf}\simeq 0.25 v_{\rm vir}$ also matches the characteristic post-shock value $v_{\rm post}\simeq \{0.25+ \mathcal{O}({\mathscr M}^{-2}) \}v_{\rm vir}$.  
Note that this hot gas coexists with relatively cool gas with $T\simeq 1000-2000 \ {\rm K}$. As noted in Section~\ref{subsubsec:pen1}, the absence of the gas with $T \lesssim 1000 \ {\rm K}$ is due to our omission of H$_2$ self-shielding against the photodissociation by LW background radiation. 
The actual gas thermal states below the ZoNR, or $T \lesssim 5000 \ {\rm K}$, 
 depend on the photodissociation efficiency.
However, we do not consider the realistic evolution of such low-temperature gas, as we ignore the LW radiation field from stars born in the central disc. 
Once the gas experiences shock heating to enter the ZoNR, succeeding thermal evolution is known to be insensitive to the pre-shock H$_2$ abundance \citep{InayoshiOmukai2012}. 
Therefore, our above estimate of the ZoNR gas should be less affected by the uncertainty in the evolution of the gas below the ZoNR. 



We estimate the mass of the ZoNR gas as $\sim 10^4~\Msun$ at this snapshot.
While the ZoNR gas component observed in the simulation result suggests the possible SMS formation, it is still far from conclusive, because the central disc structure should vary with stellar feedback neglected in the current work (see Section \ref{subsec:feedback} for discussion). In addition, the spatial resolutions in our SPH simulations are not necessarily sufficient to follow the gas thermal evolution in post-shock layers, where the gas cools down under a given constant pressure \citep{InayoshiOmukai2012}.
We thus evaluate a maximal amount of the ZoNR gas, applying the Rankine-Hugoniot jump conditions Eqs.~\eqref{eq:RU_u} - \eqref{eq:RU_M} to the fast component with the supersonic infall velocity $v_{\rm inf}>c_{\rm s}$ found in the simulation data.


The middle and bottom panels in Fig.~\ref{fig:rho_T} show the results of such analyses. 
The middle panels show the density and temperature of the fast component with $v_{\rm inf}>v_{\rm vir}$. We confirm that the fast component distributes outside the ZoNR. We assume that all the fast gas particles instantly experience shocks, for which their physical quantities are used as pre-shock values. The bottom panels show the post-shock densities and temperatures of the fast component calculated using the jump conditions. We see that the estimated post-shock temperatures are well above the lower boundary of the ZoNR. These should represent the temperatures immediately after the shock before Ly$\alpha$ cooling operates.
The black open circles in these panels represent the mass-weighted mean values of the density and temperature of the fast component measured at the shock radius $r_{\rm shock}$. Its pre-shock and post-shock values approximately trace the transition across the shock for the fast-component gas.  


As shown in the bottom panels in Fig.~\ref{fig:splash_others}, the gas expected to enter the ZoNR distributes along the filamentary accretion flows shown in the third row of Fig.~\ref{fig:splash_Mdot} within the disc radius $r_{\rm disc}\simeq 0.05 r_{\rm vir}$. This indicates the streams from equatorial directions play a major role in creating the dense shock at the central disc with $n_{\rm H}\gtrsim 10^4 \ \cmc$. The high-velocity stream from polar directions may also contribute to providing the ZoNR gas, especially near the disc centre within $r\lesssim 0.01 r_{\rm vir}$.

\begin{figure}
	\includegraphics[bb=0 0 300 550,width=8cm,scale=0.2]{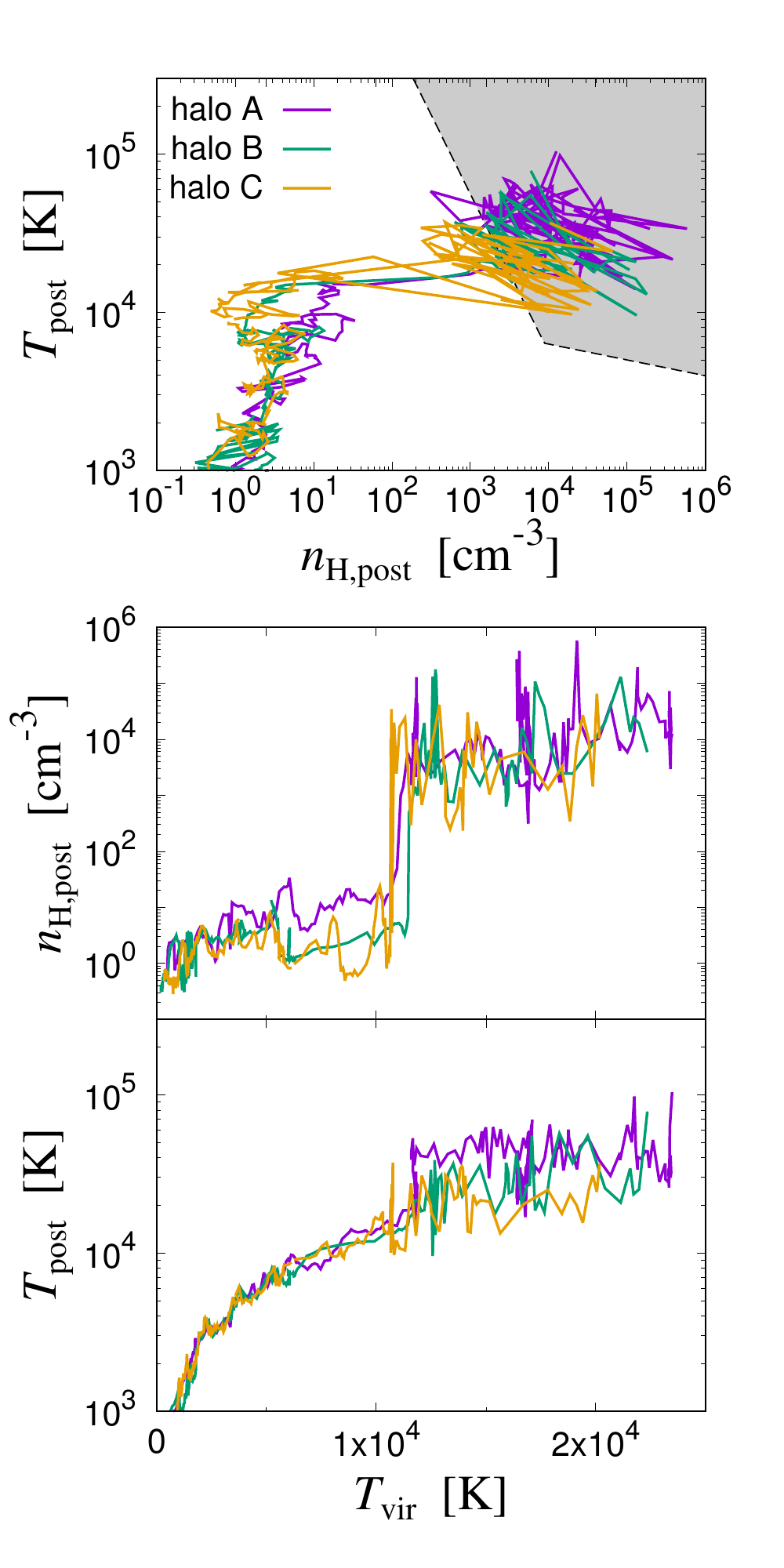}
    \caption{
Time evolution of the post-shock gas density $n_{\rm H, post}$ and temperature $T_{\rm post}$ evaluated by jump conditions. The values are calculated from the infalling velocity, density, and temperature of the fast component at the representative shock position $r_{\rm shock}$. In each panel, purple, green, and orange lines represent the different cases of halos A, B, and C, respectively. The top panel presents the evolutionary tracks on the density-temperature plane. The gray area represents the Zone of No Return, the same as in Fig.~\ref{fig:rho_T}. The middle and bottom panels show the evolution of the post-shock density and temperature against the virial temperature $T_{\rm vir}\propto M_{\rm halo}^{2/3}(1+z)$, increasing functions of time.
}
    \label{fig:post_shock}
\end{figure}

While we have looked into one snapshot at the epoch of $z=17.2$ above, we further consider the cosmological history of the possible creation of the ZoNR gas. Fig. \ref{fig:post_shock} shows the tracks of the expected post-shock density and temperature of the fast component measured at $r_{\rm shock}$ (i.e. open circles in Fig.~\ref{fig:rho_T}) in all the snapshots. The top panel shows that the post-shock gas enters the ZoNR in many snapshots. The middle and bottom panels show that the shocked gas continues to enter the ZoNR almost always, after the virial temperature exceeds $T_{\rm vir}=1.1 \times 10^4 \ {\rm K}$ at which the penetrating accretion emerges.

\begin{figure}
	\includegraphics[bb=0 0 400 280,width=9cm,scale=0.2]{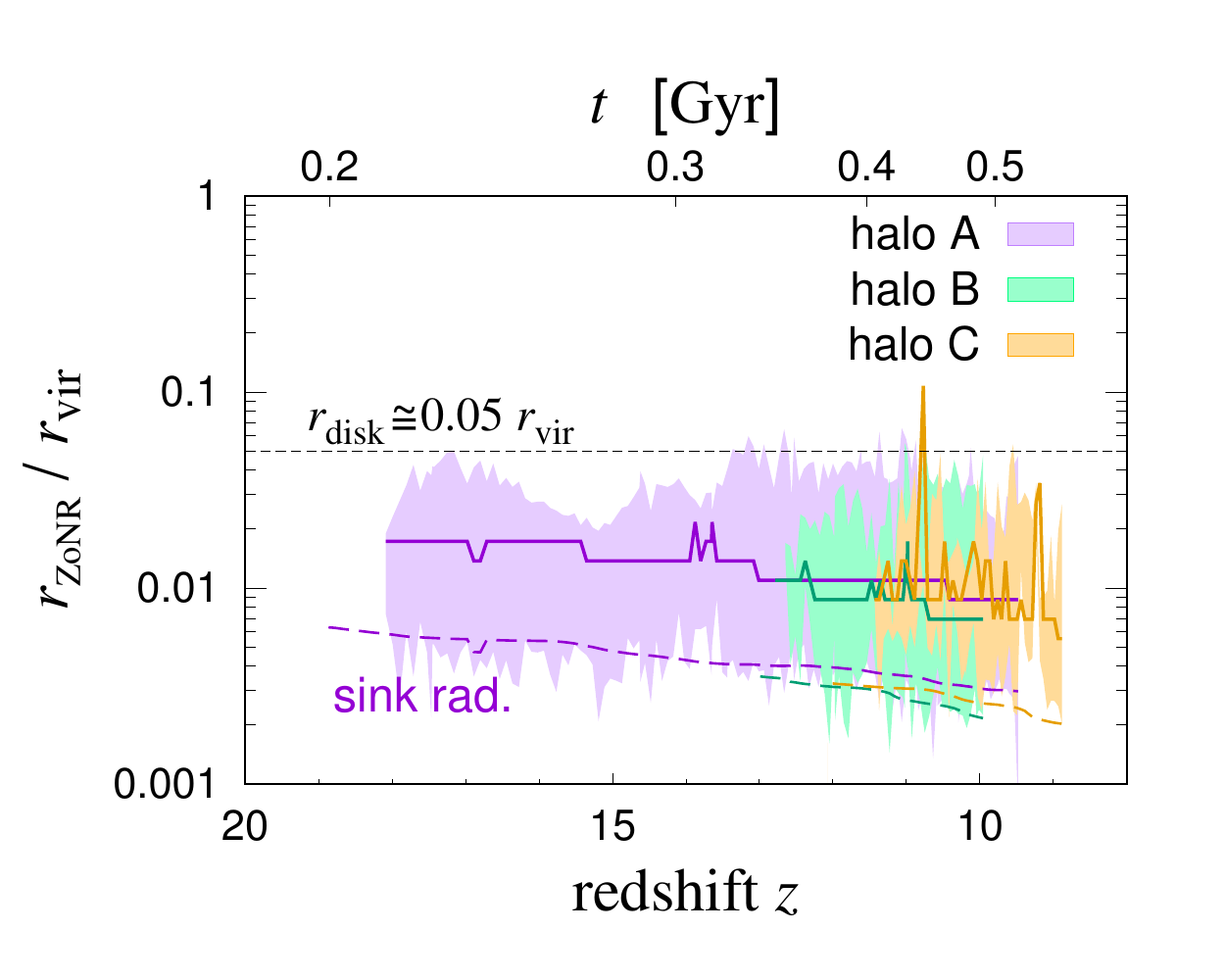}
    \caption{
Time evolution of the radial distributions of the gas particles which enter the ZoNR as post-shock states. The different colours represent the different cases of halo A (purple), B (green), and C (orange), and the hatched regions represent the radial extent of the particles. The solid lines in the middle of the hatched regions represent our estimates of the shock radius $r_{\rm shock}$. The dashed lines correspond to the sink radii measured from the halo centre. The black horizontal dashed line indicates the typical disc radius, $0.05~r_{\rm vir}$.  
}
\label{fig:R_ZoNR}
\end{figure}

Fig. \ref{fig:R_ZoNR} shows the evolution of the radial distribution of gas particles whose evaluated post-shock states enter 
the ZoNR.\footnote{
In Fig. \ref{fig:R_ZoNR}, some gas particles distribute even inside the sink radius, because we only remove them every several tens of timesteps (see also Section~\ref{subsubsec:pen2}).
} 
This figure shows that in the case of halo A, some gas always enters the ZoNR after the accretion flows come into the halo center at $z \lesssim 18$. The radial distribution range of such a component nearly corresponds to the disc radius, $r_\mathrm{disc} \sim 0.05 r_{\rm vir}$. This reinforces our argument that penetrating accretion flow eventually hits the central disc and creates shocks providing the dense and hot medium available for the SMS formation.

\subsubsection{Jeans condition for the cloud collapse}
\label{subsubsec:Jeans}

The above analysis shows that the penetrating accretion flow provides some amount of the gas which potentially enters the ZoNR. For leading to the SMS formation, a sufficient amount of such dense and hot gas needs to accumulate, exceeding the Jeans mass $M_{\rm J}$
\begin{eqnarray}
  M&>&M_{\rm J}\nonumber \\
  &=&2.7\times 10^5 \ {\rm M}_\odot
     \left(\frac{n_{\rm H}}{10^4\ {\rm cm^{-3}}}\right)^{-1/2}
     \left(\frac{T}{8000\ {\rm K}}\right)^{3/2}.
\end{eqnarray}
We here consider the mass evolution of the ZoNR gas, based on our maximal estimate using the shock jump conditions. 

\begin{figure*}
	\includegraphics[bb=0 0 780 270,width=20cm,scale=0.2]{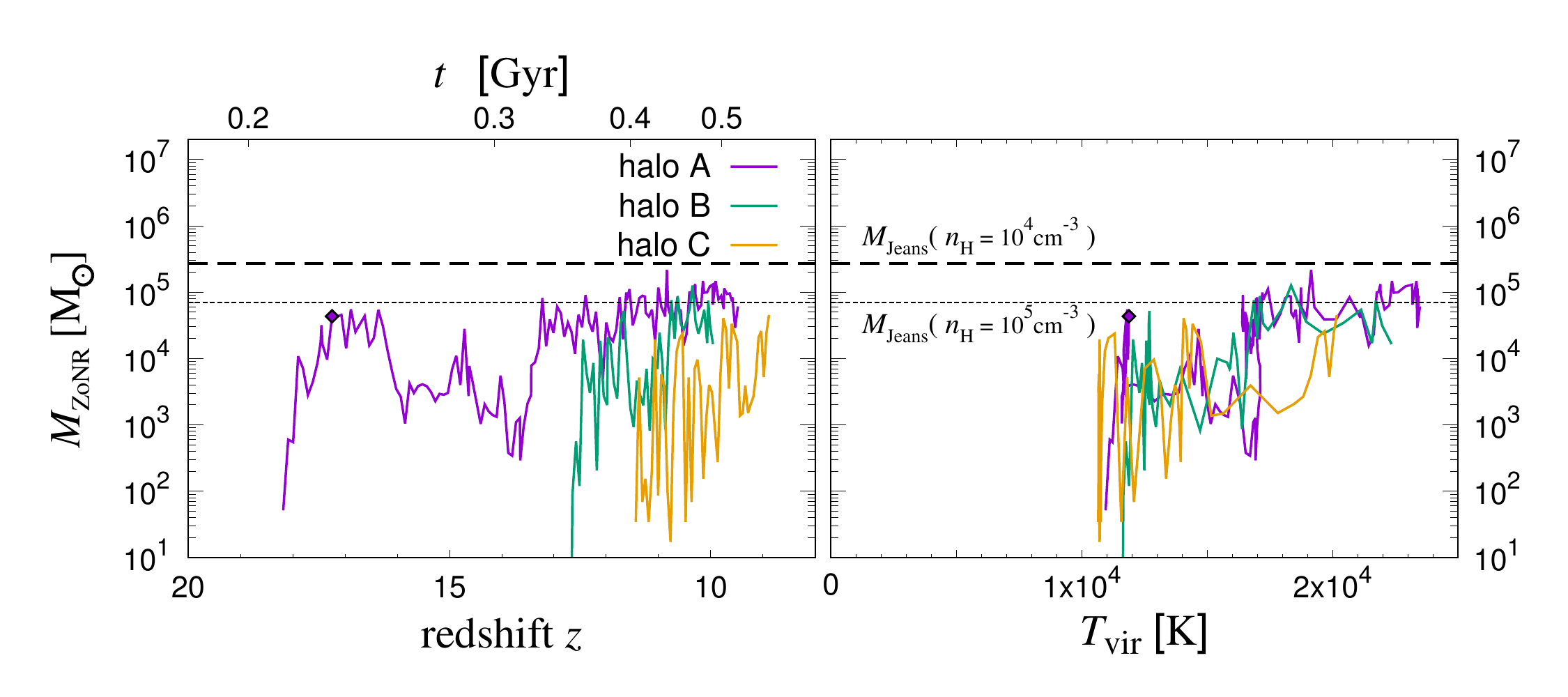}
    \caption{
Time evolution of the mass of gas entering the ZoNR as post-shock states. The left and right panels show the evolution against the redshifts (and cosmic age) and halo virial temperatures $T_{\rm vir}$. The purple, green, and orange lines represent the cases of halo A, B, and C, respectively. Note that plotted are not cumulative but instantaneous mass estimates.
In each panel, the horizontal dashed and dot-dashed lines represent the Jeans masses for different densities $n_\mathrm{H} = 10^{4-5}~\cmc$ at the given temperature $T=8000$~K. The diamonds on the purple lines denote the epoch of the snapshot presented in Figs. \ref{fig:splash_Mdot}, \ref{fig:splash_others}, and \ref{fig:rho_T}.
}
\label{fig:M_ZoNR}
\end{figure*}

Fig. \ref{fig:M_ZoNR} presents the cosmological evolution of the ZoNR mass in our simulations. The purple lines represent the case of halo A. Note that plotted are not cumulative but instantaneous mass estimates derived from the analysis of individual snapshots. This figure shows that the mass entering the ZoNR is comparable to the Jeans mass $M_{\rm J} \simeq 10^{4-5} \ \Msun$ with $n_{\rm H}=10^{4-5} \ \cmc$ and $T=8000$~K. In particular, just after the halo virial temperature exceeds $T_{\rm vir}\simeq 1.1\times 10^4$~K, $M_{\rm ZoNR}$ attains $\gtrsim 7 \times 10^4\ \Msun$. Fig.~\ref{fig:M_ZoNR} also shows $M_{\rm ZoNR}$ increases as the halo grows in mass. These suggest the possible SMS formation channel enabled by the penetrating accretion in the early universe. 

\subsection{General trends in cases of halo A, B, and C}
\label{subsec:3halos2}

In addition to the case of halo A focused in Section~\ref{subsec:haloA2}, we apply the same analyses to the other cases of halo B and C to provide a comprehensive view. Fig.~\ref{fig:post_shock} shows similar trends among these cases regarding the post-shock densities and temperatures. The inferred post-shock states almost always enter the ZoNR once the virial temperature exceeds $\sim 10^4$~K for all the cases. Fig.~\ref{fig:R_ZoNR} shows that the gas expected to enter the ZoNR always distribute within the radius of the central discs. The right panel in Fig. \ref{fig:M_ZoNR} shows that in the case of halo C, where the halo mass growth 
occurs at the lowest redshifts,
the total gas mass entering the ZoNR $M_\mathrm{ZoNR}$ is lower than those for the other cases at a given $T_\mathrm{vir}$. 
Even in this case, however, $M_\mathrm{ZoNR}$ continues to increase as the virial temperature rises.
The effect of the cosmic expansion does not prevent creating dense shocks at redshifts $z\simeq 10-20$.


In summary, our analyses suggest that the amount of gas entering the ZoNR may be sufficient for causing the gravitational collapse leading to the SMS formation.
We estimate the cosmological occurrence rate of this SMS formation channel 
in Section~\ref{subsec:Number_of_SMSs}.

\section{DISCUSSION}
\label{sec:discussion}

\subsection{Effect of sink particles}
\label{subsec:sink}

We discuss the effect of sink radius on the gas structure around sink particles since the sink radius is artificially imposed by our numerical procedure, not physically motivated.
In our prescription, the critical density for the sink creation $n_{\rm H, crit}$ is closely related to the sink radius, so we use $n_{\rm H, crit}$ as the proxy for the sink radius in this subsection.
We set the sink radius to be 10 times larger than the smoothing length $h_\text{sml}$ of the original gas particle,
and $h_\text{sml}$ decreases as the density increases.
Note that the number of SPH particles inside $h_\text{sml}$ should be constant $N_\text{neib}$ and
$h_\text{sml}$ should satisfy the following condition
\begin{eqnarray}
\frac{4}{3} \pi h_\text{sml}^3 \mu m_{\rm H} n_{\rm H, crit} = N_{\rm neib} m_{\rm part}, \nonumber \\
h_\text{sml} \propto n_{\rm H, crit}^{-1/3}m_{\rm part}^{-1/3},\nonumber
\end{eqnarray}
where $\mu$ is the mean molecular weight, $m_\text{part}$ is the mass of the gas particle.
In our fiducial case, we set $n_{\rm H, crit} = 2 \times 10^6 \ {\rm cm^{-3}}$,
resulting in the sink radius $2-3 \ {\rm pc} \simeq (0.002-0.006)r_{\rm vir}$.
Reducing the critical density by a factor of $10$ increases the sink radius by a factor of $\sim 2$.

To see how the different sink radius changes our numerical results,
we conduct the simulation with three different sink radii,
where we set $n_\text{H, crit} = 2\times 10^4$ (n04), $2\times 10^5$ (n05), and $2\times 10^6~\mathrm{cm^{-3}}$ (n06)  corresponding to that discussed in the previous sections.
In n04 and n05, we reduce the mass resolution, adopting $8$ times larger particle mass.

\begin{figure}
	\includegraphics[bb=0 0 400 250,width=9cm,scale=0.2]{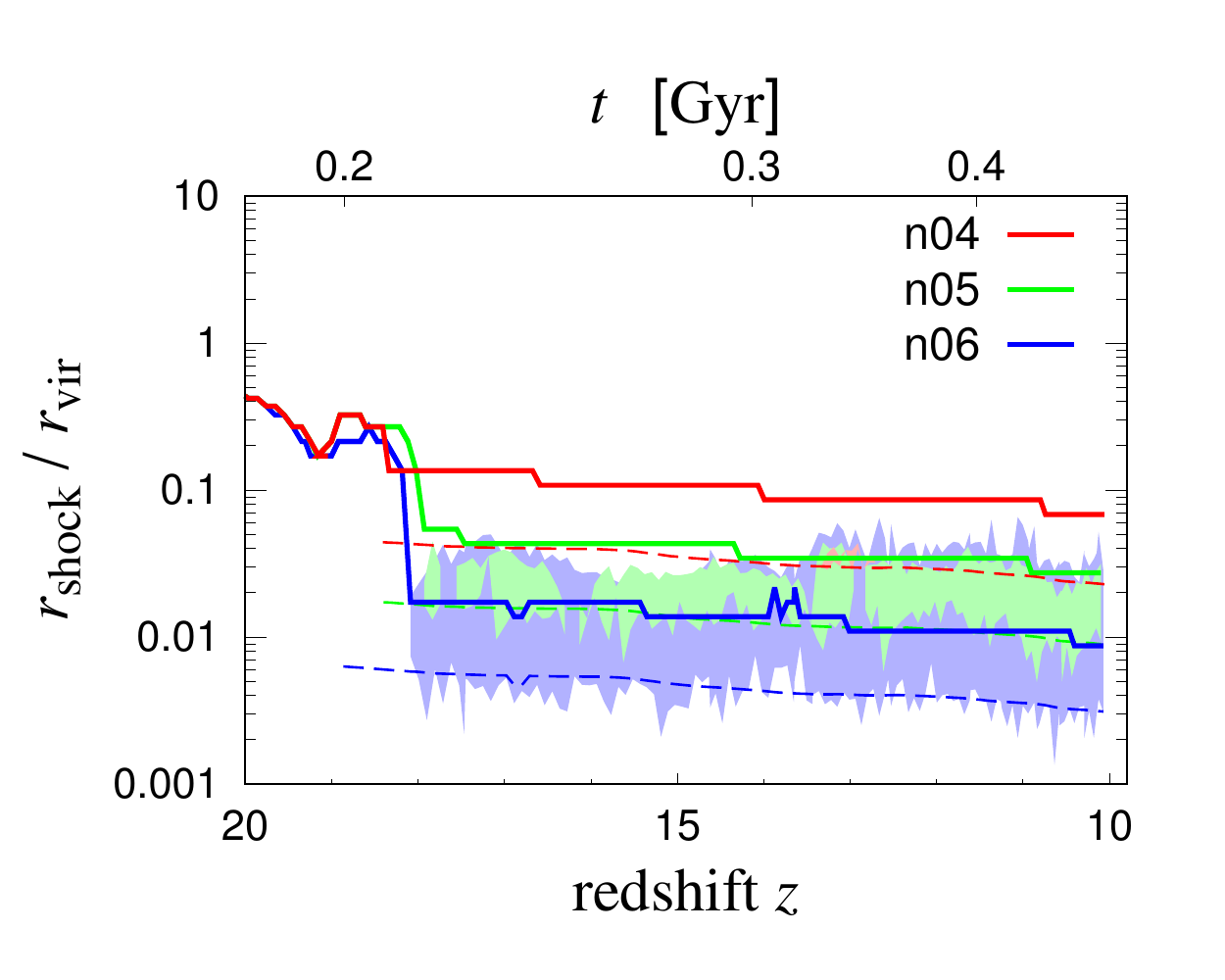}
\caption{
    The same as Fig.~\ref{fig:R_ZoNR} but for cases of halo A using different sink-creation threshold densities. The blue, light-green,  and red lines represent the cases with threshold densities of $2 \times 10^6$, $2 \times 10^5$, and $2 \times 10^4~\cmc$ (cases n06, n05, and n04), respectively.
}
    \label{fig:sinkrad_Rshock}
\end{figure}

In Fig. \ref{fig:sinkrad_Rshock}, the solid lines show the representative shock radii for different $n_\text{H, crit}$ and the dashed lines show the sink radius $r_{\rm sink}$.
Note that the shock radius follows in most time $r_{\rm shock} = 3 r_{\rm sink}$ for all three models,
that is the lowest value allowed by our definition of $r_{\rm shock}$,
indicating that the shock radius is comparable to the sink radius.
The hatched regions show the radial distribution of the gas particles, which experience shock heating and satisfy ZoNR conditions.
The cases with n05 and n06 show that this shock-heated gas is located at the distances between the sink radius and $\lesssim 0.05r_{\rm vir}$.
In the case with n05, the gas entering the ZoNR is restricted in the relatively outer region $r \gtrsim 0.01r_{\rm vir}$ compared to that in the case of n06 with $r \gtrsim 0.003r_{\rm vir}$.
In the case with n04, negligible gas is inside ZoNR since the sink size is comparable to the disc radius $\simeq 0.05 r_{\rm vir}$ and it masks out the shocked region.
This indicates that we will underestimate the mass of the shock-heated gas when we use the larger sink radius.

\begin{figure}
	\includegraphics[bb=0 0 400 270,width=9cm,scale=0.3]{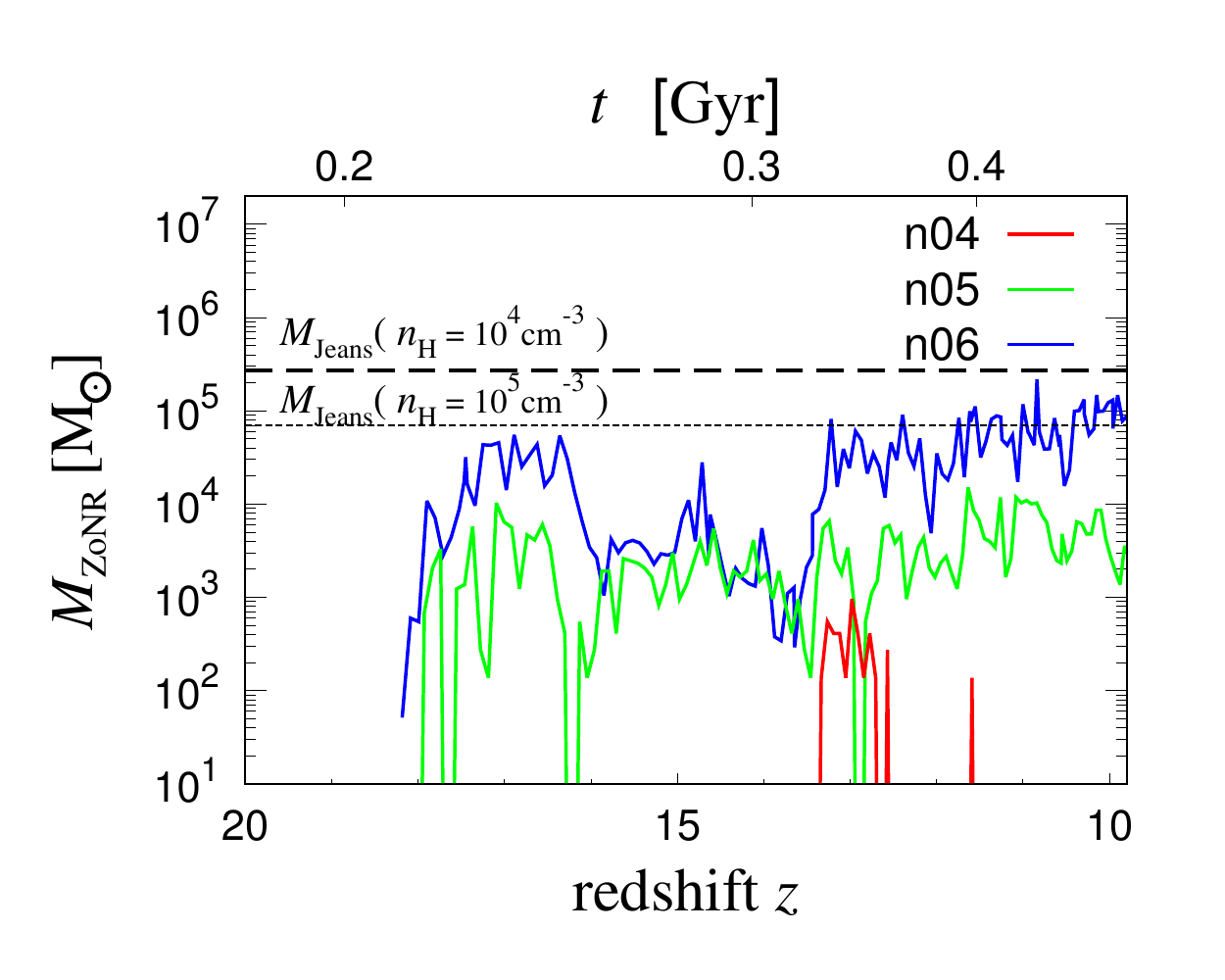}
    \caption{
The same as the left panel in Fig.~\ref{fig:M_ZoNR} but for cases of halo A using different sink-creation threshold densities. The blue, light-green, and red lines represent the cases with threshold densities of $2 \times 10^6$, $2 \times 10^5$, and $2 \times 10^4~\cmc$ (cases n06, n05, and n04), respectively.
}
    \label{fig:M_ZoNR_sinkrad}
\end{figure}

Fig. \ref{fig:M_ZoNR_sinkrad} shows the time evolution of the gas mass inside the ZoNR using post-shock density and temperature.
This demonstrates that the gas mass inside ZoNR increases as we adopt a smaller sink radius.
For example, about $10$ times larger gas mass is inside ZoNR in model n06 than in n05, while negligible gas mass enters ZoNR in n04.
We do not perform the run with smaller sink radii and higher resolution until the amount of the gas inside ZoNR converges, due to our limited computational resources.
This indicates that adopting the smaller sink radius can increase the amount of shock-heated gas,
that would be preferable for the formation of SMSs.

\subsection{Comparisons with previous studies}
\label{subsec:diff}

In Sections \ref{subsec:3cases} and \ref{subsec:interpretation}, we have presented our results in the context of previous cosmological simulations. Fig.~\ref{fig:massevo} provides a comprehensive view, suggesting that the semi-analytic framework by \citet{BirnboimDekel2003} is also applicable to the emergence of the cold accretion in the early universe. However, it is not straightforward to make comparisons with previous simulations using different numerical techniques and analysis methodologies. We discuss the relevant points here in more detail. 


\subsubsection{\citet{WiseAbel2007}}

For instance, \citet{WiseAbel2007} study the detailed gas dynamics through the virialization of ACHs using N-body + AMR code {\tt ENZO} \citep{Bryan2014}. They demonstrate that cosmological accretion flow enters deep inside a halo by the effects of radiative cooling, comparing different runs where they artificially control the cooling processes. The cold flows reach the radius $r \simeq r_{\rm vir}/4$ at their final snapshot, shortly after a cloud collapse occurs via hydrogen atomic cooling near the halo centre. The mass of the halo at the emergence of the cold accretion is slightly higher than our derived critical masses as illustrated in Fig.~\ref{fig:massevo}.


Let us count the differences between our work and \citet{WiseAbel2007}. A difference is in the treatment of chemistry and resulting cooling processes. We have used moderate LW background radiation to destroy H$_2$ molecules and prevent normal Pop I\hspace{-1pt}I\hspace{-1pt}I star formation in mini-halos (Section~\ref{sec:methods}). 
During the cloud collapse in the ACHs, however, H$_2$ molecules form and operate as a coolant (Fig.~\ref{fig:rho_T_Fer}). H$_2$ molecular cooling becomes effective for the central gas disc forming afterward. 
In contrast, \citet{WiseAbel2007} for simplicity ignore H$_2$ molecular cooling for the case where they examine the development of cold accretion in ACHs. Such differences may alter the gas structure near the halo centres and the dynamics of the accretion flow therein. 


A more striking difference is in the numerical methods for solving the gas dynamics: SPH in our case and AMR in \citet{WiseAbel2007}. 
Some previous studies show that the gas dynamics at the halo virialization depends on the numerical methods. For instance, \citet{Nelson+2013} study thermal properties of the accretion flow toward $\sim 10^{10-12}~\Msun$ halos at $z=2$ observed in their simulations using different codes. 
Whereas these halos differ from ours in the mass and epoch, they demonstrate that the SPH simulations using {\tt GADGET-3} code tend to overestimate the significance of the cold accretion within $\sim 0.5 r_{\rm vir}$ compared to runs using moving-mesh code {\tt AREPO} \citep{SpringelArepo2010}. \citet{WiseAbel2007} report the presence of turbulence stronger than ours in the deep interior of the ACHs, which may be caused by fluid instabilities not captured by the standard implementation of SPH \citep[e.g.][]{Agertz2007}. 


Another difference is in the analysis methodology to evaluate how deeply the accretion flow penetrates the ACHs. As outlined in Appendix~\ref{sec:shockposition}, we have made use of particle distributions on the radial position-velocity maps to evaluate the typical shock radius $r_\mathrm{shock}$ (see also Fig.~\ref{fig:r-v}). \citet{WiseAbel2007}, in contrast, use 2D slicing maps of the adiabatic invariant $K=T/n^3$ for that purpose (their Fig.~3). We did not rely on the adiabatic invariant because any gas components suffer from very efficient Ly$\alpha$ cooling near the halo centres (Section~\ref{subsubsec:pen1}). With this method, it is difficult to extract only the components with large radial velocities deep in the ACH, if any. If supersonic turbulence remains in the deep interior of the ACHs, however, our method is ineffective in finding coherent accretion streams. 


\subsubsection{\citet{Greif2008}}

\citet{Greif2008} study the assembly of the first galaxy in an ACH, following the Pop I\hspace{-1pt}I\hspace{-1pt}I star formation in ancestral mini-halos and mass accretion onto the resulting BHs.
They perform cosmological zoom-in simulations with their N-body + SPH code.
They report that the hot accretion mode is dominant in a mini-halo with $M_{\rm halo}=6 \times 10^5 \ \Msun$ at $z=23$, while the cold accretion mode is dominant in the ACH with $M_{\rm halo} = 5\times 10^7 \ \Msun$ at $z=10.62$. The cold flows come into the centre of the ACH at this epoch (see their Figs. 8 and 10), which is consistent with our simulation results and semi-analytic minimum halo mass.
Note that they set no LW background radiation, which somewhat facilitates the penetration of cold flows by efficient ${\rm H}_2$ cooling.


\subsubsection{\citet{Fernandez2014}}

\citet{Fernandez2014} study whether the SMSs form by the cold accretion in ACHs as proposed by \citet{InayoshiOmukai2012}.
They perform cosmological zoom-in simulations using N-body + AMR code {\tt ENZO} \citep{Bryan2014}. We have followed their method of assuming moderate LW background with $J_{21}=10$ to suppress ${\rm H_2}$ cooling and resulting Pop I\hspace{-1pt}I\hspace{-1pt}I star formation in mini-halos. \citet{Fernandez2014} do not report the emergence of the cold accretion, which is consistent with our "minimum halo mass" line on the $z-M_{\rm halo}$ plane as demonstrated in Fig.~\ref{fig:massevo}. Whereas they terminated the simulations shortly after the cloud collapse in the ACHs, we followed the evolution afterward using sink particles.
Other differences include numerical methods for solving hydrodynamics and the criterion of where the shock is created.
As noted above, SPH simulations we use tend to overestimate the significance of cold accretion, compared to AMR.
On the criterion of shock position, they use slicing maps of gas entropy, Mach number, and velocity divergence $(-\nabla \cdot {\bf v})$.

\subsubsection{\citet{Latif2022}}

\citet{Latif2022} investigate the possibility of SMS formation in a very rare case where an ACH forms at the intersection of cold filaments at $z\simeq 25$, performing cosmological simulations using the {\tt ENZO} code \citep{Bryan2014}.
They report that the Pop I\hspace{-1pt}I\hspace{-1pt}I star formation is prevented for $4 \times 10^5 \ \Msun \lesssim M_{\rm halo} \lesssim 4\times 10^7 \ \Msun$ due to strong turbulent pressure caused by the cold flows, rather than the suppression of ${\rm H}_2$ formation by LW radiation or other processes previously considered. The cold flows eventually dominate the turbulent pressure and create a cloud at the halo centre.
They also follow the long-term evolution after the emergence of the cold accretion, demonstrating that multiple SMSs form in the cloud. 


Regarding the penetration of the cold flows, their Fig.~1 
indicates that the flows penetrate the halo by the epoch of $z \simeq 25$. They assume no LW background radiation, allowing ${\rm H}_2$ formation in relatively diffuse gas, as in \citet{Greif2008}.
As shown in Fig.~\ref{fig:massevo}, their result is more or less consistent with our simulation results and semi-analytic minimum halo mass.
This may be surprising because they study the rare case found in dozens of 37.5~Mpc cosmological boxes with different initializations, while we have chosen more typical ACHs.  


The role of the cold accretion in the SMS formation differs from what we suppose. 
During the cloud collapse in their simulation, H$_2$  molecular cooling is effective, and the gas temperature is $\lesssim 10^3$~K. The corresponding thermal evolution track goes below the ZoNR on the density-temperature plane. Nonetheless, very rapid accretion onto protostars at the rate $\sim 0.1-1~\mdotsunyr$ occurs owing to the dynamics provided by large-scale cold accretion. 
They report that any coherent structure such as the central disc is destroyed by strong turbulence. 
Such evolution is not found in our simulations, which may be attributed to the rareness of the case studied in \citet{Latif2022}.

\subsection{Number density of SMSs}
\label{subsec:Number_of_SMSs}


In Section~\ref{sec:results}, we have considered ACHs at $10 \lesssim z \lesssim 20$ to show that the SMS formation can be induced by dense shocks provided by cold accretion. 
The ACHs we chose corresponds to $\simeq 2 \sigma$ peak, whose number density is $n_{\rm halo} \sim 1~{\rm cMpc^{-3}}$. In contrast, SMBHs observed at $z>6$ are rare objects of $\sim 1-10~{\rm cGpc^{-3}}$. If all halos similar to ours host SMSs, it will overproduce the massive seed BHs. It is thus reasonable that SMS formation needs some further conditions in reality. We here discuss the actual occurrence rate of the SMS formation, considering further additional conditions. 


\citet{LiInayoshi2021} show that the heavy seed BHs provided by the SMSs that eventually grow into SMBHs at $z\simeq 6-7$ should have formed at the epoch of $z \simeq 30$. We therefore consider the ACHs at $z=30$, whose number density is $n_{\rm ACH} \simeq ({\rm d}n/{\rm d}\log M_{\rm halo})\sim 10^{-3}~{\rm c Mpc^{-3}}$ \citep{BarkanaLoeb2001}. We also require that progenitor halos of the ACHs have not undergone Pop III star formation to avoid metal enrichment. \citet{Fernandez2014} evaluate such metal-free halo fractions as $f_{\rm p}\simeq 2\times 10^{-4}$ at $z=10$ with DM halo merger trees, assuming moderate LW background $J_{\rm 21}=10$. We adopt this value for ACHs at $z=30$ as a rough estimate. 
In Section~\ref{subsubsec:pen1}, we have shown that the accretion stream reaches the halo centre $\Delta t \sim 10$~Myr after the cloud collapse first occurs within the ACHs. However, the shorter time lag $\Delta t$ is favored for the SMS formation because in-situ star formation prior to the intrusion of the cold stream should also induce the metal enrichment. The more rapid development of the cold accretion is realized with merger events, possibly those with $\gtrsim 10^{6}~\Msun$ halos to provide a gas cloud with $\gtrsim 10^5~\Msun$. Recall that the gas accretion rate at $r=r_{\rm disc}$ is comparable to that measured at $r=r_{\rm vir}$ (Fig.~\ref{fig:r-Mdot}). Extrapolating the halo merger rates given by \citet{Fakhouri+2010} to low-mass ranges, we evaluate the frequency with which a given halo with $M_{\rm halo} \sim 10^7~\Msun$ merges with halos with $M_{\rm halo}' \gtrsim 10^6~\Msun$ per unit time at $z=30$ as
\begin{equation}
    \frac{{\rm d}N_{\rm merger}}{{\rm d}t} \simeq 2\times 10^{-3}~{\rm Myr^{-1}}
    \left(\frac{1+z}{31}\right)^{2.6}
    \left(\frac{M_{\rm halo}}{10^7~{\rm M_\odot}}\right)^{1.13}
    \left(\frac{M_{\rm halo}'}{10^6~{\rm M_\odot}}\right)^{-1}.
\end{equation}
The number density of SMSs or halos fulfilling all the above conditions is estimated as 
\begin{eqnarray}
n_{\rm SMS} &=&n_{\rm ACH}~f_{\rm p}~\frac{{\rm d}N_{\rm merger}}{{\rm d}t}\Delta t\nonumber \\
&\sim&10^{-9}~{\rm cMpc^{-3}} 
\left(\frac{\Delta t}{3~{\rm Myr}}\right),
\label{eq:nsms}
\end{eqnarray}
for which we use $\Delta t=3~{\rm Myr}$, the typical lifetime of massive Pop III stars.
Eq.~\eqref{eq:nsms} indicates that the seed BH number density realized by the considered channel can be comparable to that of SMBHs exceeding $10^9~\Msun$ observed at $z > 6$. We also estimate $n_{\rm SMS} \sim 2\times 10^{-7}~{\rm cMpc^{-3}}$ at $z=20$, and $2\times 10^{-6}~{\rm cMpc^{-3}}$ at $z=10$, respectively. These seed BHs might evolve into SMBHs with $M_{\rm BH}\sim 10^{7-8}~\Msun$ powering relatively fainter quasars.


Note that the above is a rough estimate with some uncertainties.
For example, \citet{Fernandez2014} show that the metal-free fraction of ACHs $f_{\rm p}$ becomes larger than our adopted value by more than an order of magnitude with a slightly stronger LW background $J_{\rm 21}=30$ at $z=10$. 
It is uncertain how such a dependence changes with increasing redshifts. Our choice of the time duration $\Delta t = 3~{\rm Myr}$ is also arbitrary, but it is conservative to avoid any supernova and metal enrichment in the ACH. 
Some studies suggest that the SMS formation is possible even under metal enrichment up to $\sim 10^{-3}~Z_\odot$ \citep{InayoshiOmukai2012, Chon+2020}.
Since the metal enrichment by a single SN is $\lesssim 10^{-3}~Z_\odot$ \citep[e.g.][]{Wise_Turk2012, Ricotti+2014}, in such a case several SNe may be allowed for SMS formation (see also Section~\ref{subsubsec:SN}). The corresponding $\Delta t$ should be $\sim 10-100~{\rm Myr}$, which multiplies our SMS number density by $\sim 3-30$.
We also note that our estimate is based on the simulations without stellar radiative feedback, which can alter the gas density and temperature, the morphology of gas clouds, and possibly the behavior of cold accretion. We discuss this point in Section~\ref{subsubsec:rad}.




\subsection{Caveats: feedback effects}
\label{subsec:feedback}

Since we have followed the long-term ($\sim 0.1$~Gyr) evolutions after the first collapse of a gas cloud in our simulations, the subsequent star formation should cause feedback via different channels in reality. We here discuss the possible roles of such feedback effects, which we have ignored for simplicity.

\subsubsection{UV radiative feedback}
\label{subsubsec:rad}

We first consider the effect of the stellar UV feedback. Since our simulations show that the emergence of the cold accretion occurs later than the cloud collapse, which leads to the star formation, the stellar UV feedback may affect the subsequent evolution. Stellar UV photons heat the gas up to $T \sim 10^4$~K, and the ACH virial temperature is slightly lower than that. The photoheating effect thus may induce the photoevaporation of the halo gas. 
\citet{Pawlik+2013} study the stellar UV photoheating effect during the assembly of first galaxies performing cosmological simulations. They show that the photoheating hardly changes the galaxy structure once the halo virial temperature exceeds $\sim 10^4$~K. Moreover, the filamentary accretion flow tends to be protected against the UV photons coming from sources within the halo \citep[e.g.][]{Pawlik+2013,Chon+2017}. We thus naively expect that the cold accretion should start for $T_\mathrm{vir} \gtrsim 10^4$~K even under the stellar UV photoheating effect, albeit with some delay. 


The dense gas disc forming near the halo centre (see Figs.~\ref{fig:splash_Mdot} and \ref{fig:splash_others}) is the possible site for the subsequent star formation. Their UV feedback also affects the structure of the central disc. Regarding the SMS formation, there are two competing UV feedback effects on hydrogen chemistry: photoionization and photodissociation. The former promotes H$_2$ formation by supplying electrons available as catalysts of the H$^-$ channel, and the latter counteracts by destroying ${\rm H_2}$ molecules.
\citet{InayoshiOmukai2012} argue that the evolution of gas clouds is insensitive to the initial chemical fraction $x_{\rm e}$ and $x_{\rm H_2}$ as long as they are once heated by dense shock up to $T\gtrsim 5000 \ {\rm K}$ with $n_{\rm H}\gtrsim 10^4 \ \cmc$.
This is because efficient collisional dissociation resets even a high H$_2$ fraction to much lower values and $x_{\rm e}$ decreases as the gas cools. 
Therefore, once the gas is shock heated to enter the ZoNR, the subsequent thermal evolution is insensitive to
the chemical abundance $x_{\rm e}$ and $x_{\rm H_2}$ in a pre-shock state. Even in this case, photoionization and photodissociation may change the gas thermal evolution if the post-shock state quits the ZoNR. This potentially occurs when the post-shock medium takes too long time to assemble enough for the gravitational collapse, $M_{\rm ZoNR}>M_{\rm J}$, within the ZoNR.
In this case, formation of H$_2$ molecules starts to overcome the H$_2$ collisional dissociation, and resulting H$_2$ molecular emission cools the gas down to a few $\times$ 100~K without additional effects. The radiative feedback from nearby stars should modify the evolution by two competing effects of photoionization and photodissociation. Simulations by \citet{Pawlik+2013} suggest that the former overcomes the latter, resulting in efficient H$_2$ formation up to $x_{\rm H_2}\sim 10^{-4}-10^{-3}$, though they follow only relatively low-density gas with $n_{\rm H} < 10^2 \cmc$.


\subsubsection{Supernova feedback and metal enrichment}\label{subsubsec:SN}

We next discuss the effect of supernova (SN) feedback and resulting metal enrichment. In Section~\ref{subsec:3cases}, we have shown that the cold accretion emerges $\sim 10-30$~Myr after the first event of the cloud collapse in ACHs. SN feedback thus begins to operate early in the evolution we follow. 
Although SN feedback delays the development of the accretion stream near the halo centre, its impact should be alleviated as the halo mass increases. The ejected gas returns to the halo after a while. Updating the minimum halo mass for the cold accretion (e.g. Fig.~\ref{fig:massevo}) under UV and SN feedback is a task for future studies.


SN metal enrichment is important in terms of possible SMS formation. The critical metallicity, above which metal line cooling prevents the nearly isothermal collapse even in the ZoNR, is known as $Z_{\rm crit} \sim 10^{-3} \ {\rm Z}_\odot$ \citep{Omukai2008, InayoshiOmukai2012,Chon+2020}. Previous cosmological simulations have investigated the SN feedback and metal enrichment during and after the first galaxy formation  \citep{Wise_Turk2012, Graziani+2015, Graziani+2017, Graziani+2020, Ricotti+2014, Jeon+2017, Yajima+2017, Abe+2021}.
For instance, \citet{Wise_Turk2012} show that even a single event of pair-instability SN or hypernova enriches its $\lesssim 10 \ {\rm kpc}$ neighborhood up to $Z \sim 10^{-4} - 10^{-3} \ {\rm Z}_\odot$. \citet{Ricotti+2014} show that the metallicity remains relatively low $Z \lesssim 10^{-3} \ {\rm Z}_\odot$ only in an early phase of the first galaxy formation for $\sim 100$~Myr. These suggest that the initial $\sim 100$~Myr since the first emergence of the cold accretion may be the possible duration of the SMS formation with $Z \leq 10^{-3} \ {\rm Z}_\odot$.



\subsection{Expected events other than supermassive star formation}
\label{sec:impact}
In Section \ref{subsec:feedback}, we have concluded that the filamentary flows will survive and possibly hit the central disc even in the presence of several feedback processes. It is interesting to consider whether the shock heating triggers the SMS formation and what happens otherwise. There are two necessary conditions for the SMS formation as discussed below.

One is that the cloud should remain metal-poor to avoid fine-structure line cooling, which reduces the gas temperature below $5000~$K. When the successive SNe enrich the cloud to the level of $Z/Z_\odot \gtrsim 10^{-3}$, it operates to reduce the gas temperature below $200~$K \citep{Bromm+2001, Omukai2008}.
In that situation, rapid cooling induces vigorous fragmentation, and the massive star cluster forms instead of a SMS. 
The properties of the star clusters depend on their metallicity.
\citet{Mandelker+2018} expect that the globular cluster will form for $Z \gtrsim 0.01~Z_\odot$, considering relatively massive halos with $M_{\rm halo}\gtrsim 10^{9}~\Msun$ at redshifts $z \lesssim 8$. 
They analyse the Jeans instability of the cold accretion and find that it operates to form globular clusters in the accretion flow penetrating massive halos with $M_{\rm halo} \sim 10^{10} \ \Msun$.
\citet{Chon+2020} have found that, for the mildly metal-enriched cases with $Z/Z_\odot = 10^{-3}$, the star cluster along with central very massive stars with $100-10^3~\Msun$ form. They do not consider the large ram pressure driven by cold accretion, which would make the stellar cluster more compact.
This can induce the run-away collision between stars, and the mass of the central stars will become larger \citep[e.g.][]{Sakurai+2017}.

The other condition is that the shocked medium should be massive enough to trigger gravitational instability. 
If some feedback effects reduce the mass of the cold accretion flow, it becomes difficult to satisfy this condition. 
In this case, reducing the Jeans mass is one possible way to form massive stars by cold accretion. 
If the shock-heating operates at a high-density region with $n \gtrsim 10^6~\mathrm{cm^{-3}}$, the Jeans mass is $10^4~\Msun$ on the atomic-cooling path.
Our numerical simulations show that the shock heating occurs at a higher-density region as we increase the spatial resolution.
If H$_2$ cooling or fine-structure line cooling operates and reduces the cloud temperature, the Jeans mass becomes smaller accordingly.
The actual gas temperature depends on the balance between the heating by the feedback and the several cooling processes and also on the electron abundance.
For instance, when the cloud temperature is $\sim 1000~$K, the resulting Jeans mass is $10^4~\Msun$ at $n_{\rm H} \sim 10^5 \ \cmc$.
In both cases where the shocked region has a higher density or lower temperature, the collapse of the cloud with $10^4~\Msun$ is expected, 
possibly leading to the formation of a $\sim 10^4~\Msun$ star.
However, cooling can induce vigorous fragmentation and would result in star cluster formation instead.

To investigate the final fate of the shock-heated region by the cold accretion under the stellar feedback, we need realistic cosmological simulations that address both the sub-pc physics of individual star formation and the kpc physics of large-scale gas inflows.

\subsection{Angular momentum barrier in SMS formation}
 \ 

 The angular momentum of clouds is a potential barrier to preventing the SMS formation, as it can lead to efficient fragmentation rather than monolithic collapse. Although our simulations do not directly follow the SMS formation, we here estimate the angular momentum of a cloud formed by the cold accretion. 
We consider a cloud forming at $r=r_{\rm disc}=0.05 r_{\rm vir}$ as shown in Section~\ref{sec:SMS_pre}. Assuming that the typical rotational velocity of the gas at $r_{\rm disc}$ is comparable to the virial velocity $v_{\rm vir}$, which is the case in our simulations, the specific angular momentum with respect to the disc centre is estimated as 
\begin{eqnarray}
    l \sim 0.05 r_{\rm vir} v_{\rm vir} =6.7\times 10^{25}~{\rm cm^2 s^{-1}}
    \left(\frac{1+z}{15}\right)^{-1/2}
    \left(\frac{M_{\rm halo}}{10^7~{\rm M_\odot}}\right)^{2/3}.
\end{eqnarray}
These values of $l\lesssim 10^{25-26}~{\rm cm^2 s^{-1}}$ are typical for ACHs, for which previous studies have supposed the SMS formation \citep{Becerra+2018}.




This intrinsic angular momentum must be removed from most of the cloud gas for the SMS formation within the typical timescale of $ \sim 1~{\rm Myr}$. Several processes have been proposed as efficient transport mechanisms, such as the gravitational torque enhanced by the spiral/bar mode instability \citep{Sakurai_Vorobyov+2016, Becerra+2018, Matsukoba+2021}, that exerted from the non-axisymmetric DM host halo \citep{Chon+2016, Shlosman+2016}, and magnetic braking \citep{Pandey+2019, Haemmerle+2019}.
Indeed, \citet{Becerra+2018} show in their simulations of the SMS formation that the angular momentum of $l \simeq 10^{25}~{\rm cm^2 s^{-1}}$ at $r\simeq 10~{\rm pc}$ is efficiently transported by gravitational torque. We expect that similar processes should also work in our cases to alleviate the angular momentum barrier, which is to be verified in future numerical simulations.


\subsection{Further growth of seed Black Holes}

For explaining $\sim 10^9\ \Msun$ SMBHs at $z \gtrsim 6$, seed BHs must grow to become that massive within several $\times \ 100 \ {\rm Myr}$.
Some previous studies point out that the cold accretion provides the gas supply from the large-scale structure to the halo centre, enhancing accretion rates onto seed BHs \citep{DiMatteo2012, Smidt+2018}. Note that other studies find the opposite results that the seed BHs with $M_{\rm BH}=10^5 \ \Msun$ located at the centre of ACHs hardly grow 
because photoheating from nearby stars and BH accretion discs lowers the density of the surrounding medium \citep{Johnson+2011, Latif+2018}.
Although still controversial, whether the efficient growth of seed BHs with the cold accretion is reasonable seems to depend on the boosting factor $\alpha=\dot{M}/\dot{M}_{\rm Bondi}$, where $\dot{M}_{\rm Bondi}$ is the Bondi accretion rate. Some cosmological simulations assume $\alpha=100$ without resolving the flow structure within the Bondi radius \citep{SpringelDiMatteo2005, DiMatteo+2005, DiMatteo+2008, Sijacki+2007, Booth+2009}.


In the regime of massive halos exceeding $\sim 10^{13} \ \Msun$, small-scale physics may realize the efficient accretion corresponding to the boosting factor $\alpha \sim 100$ \citep{Gaspari+2013, Gaspari+2015, Gaspari+2017}.
However, it is uncertain whether the same is applicable to the seed BHs in less massive halos with $M_{\rm halo} \sim  10^{7-8} \ \Msun$. If the cold accretion enables the SMS formation, it is advantageous to supply seed BHs near the halo centre toward which the cold accretion streams converge. Further mass growth of the seed BHs has to be investigated by future high-resolution simulations.

\section{CONCLUSIONS}
\label{sec:conclusions}

We have studied the first emergence of the cold accretion, or the supersonic accretion flows directly coming into the halo centre, performing a suite of cosmological N-body + SPH simulations. Using the zoom-in technique, we have achieved sufficiently high spatial resolutions to study the detailed flow structure within halos with $M_{\rm halo} \sim 10^{7-8}~\Msun$ at the epochs of $z \simeq 10-20$.
We have further considered the possible SMS formation from the shocked dense gas created by the accretion flow near the halo centres \citep[][]{InayoshiOmukai2012}. To this end, we have also followed long-term evolution after the emergence of the penetrating accretion for a few $\times~100$~Myr. We make use of sink particles representing accreting Pop I\hspace{-1pt}I\hspace{-1pt}I stars to save computational costs. Our findings are summarized as follows.


Our examined three cases show that the accretion flow penetrates deep inside halos after certain epochs. Accordingly, the typical positions of the accretion shocks shift inward from $r_{\rm shock} \sim 0.1~r_\mathrm{vir}$ to $ 0.01~r_\mathrm{vir}$. Such transitions approximately occur when the halo mass exceeds 
\begin{eqnarray}
M_{\rm halo, \ min}\simeq 2.2\times 10^7 \ \Msun
\left(\frac{1+z}{15}\right)^{-3/2}, 
\label{eq:mhalomin}
\end{eqnarray}
which corresponds to the virial temperature of $T_{\rm vir} \simeq 1.1 \times 10^4 \ {\rm K}$ (Fig.~\ref{fig:z-Rshock}). This is the {\it minimum} halo masses above which the cold accretion emerges, in contrast to the {\it maximum} halo masses of $M_{\rm halo, \ max} \sim 10^{11-12} \ \Msun$ provided by previous studies supposing the massive galaxy formation at lower redshifts \citep[][]{BirnboimDekel2003}.
Each run of our simulations shows that the cold accretion emerges shortly after the first run-away collapse of a cloud in an ACH, after time-lag of $\Delta t\simeq 10-30\ {\rm Myr}$. This suggests that the emergence of the cold accretion follows the birth of normal Pop I\hspace{-1pt}I\hspace{-1pt}I stars in ACHs. 


To interpret our and previous simulation results, we have applied the semi-analytic models of the spherical accretion developed by \citet[][]{BirnboimDekel2003} to our cases of the ACHs (Fig.~\ref{fig:massevo}). Whereas the models also provide the minimum halo masses above which the cold accretion should appear, we show that the models modified to include the effects of the filamentary accretion give those in agreement with Equation \eqref{eq:mhalomin}.


The supersonic accretion flow continues until it hits the gas discs at the halo centres. The typical size of the disc is $\sim 0.05~r_\mathrm{vir}$, within which the dense accretion shock appears. As a result, the shocked dense gas accumulates on the surface layers of the discs. 
Note that most of the gas in an ACH has almost the same temperatures at $T \sim 10^4 \ {\rm K}$, regardless of the supersonic or subsonic components, owing to very efficient Ly$\alpha$ cooling. The penetrating accretion flow is not very cold relative to the surrounding medium. 

To study the possibility of the SMS formation, we have further analysed the gas dynamics in the vicinity of the central disc in detail. 
A simulation snapshot shows that there is the subsonic, dense, and hot ($n_{\rm H} \gtrsim 10^4~\cmc$ and $T \simeq 8000$~K) medium available for the SMS formation near the central disc (Fig.~\ref{fig:rho_T}).
The total mass of such "ZoNR" gas is $\sim 10^4~\Msun$. Because of our limited spatial resolution and ignorance of feedback effects, however, we did not solely rely on the actual simulation results. We have instead provided maximal estimates of the ZoNR gas, applying the shock jump conditions for gas particles that have supersonic radial infall velocities at each snapshot. 
This post-process analysis method is free from the above limitations in following the dynamics of the dense shocked medium in the simulations. 


Our analyses show that, after the emergence of the penetrating accretion, there is almost always some gas that potentially enters the ZoNR. Such gas spatially distributes in the innermost part of the halos, nearly within the size of the central disc (Fig.~\ref{fig:R_ZoNR}). These support our argument that penetrating accretion flow eventually hits the central disc and creates shocks providing the ZoNR medium available for the SMS formation. The mass of the ZoNR gas estimated by our method is $M_{\rm ZoNR} \sim 10^{4-5}\ \Msun$ in some snapshots, comparable to the Jeans mass at the densities $n_{\rm H} \sim 10^{4-5}~\cmc$ and temperature $T \simeq 8000$~K (Fig.~\ref{fig:M_ZoNR}). This means that the ZoNR gas can become gravitationally bound and ready to start the collapse. 


In this work, we have ignored processes that affect the long-term evolution after the emergence of the cold accretion, such as radiative feedback from stars, supernova feedback, and metal enrichment, for simplicity. The first emergence of the penetrating accretion and resulting SMS formation under these additional effects are intriguing and open for future studies.

\section*{ACKNOWLEDGEMENTS}

The authors express their cordial gratitude to Prof. Takahiro Tanaka for his continuous interest and encouragement. 
We sincerely appreciate Kazuyuki Omukai, Naoki Yoshida, Kohei Inayoshi, Kotaro Kyutoku, Kazuyuki Sugimura, Shingo Hirano, Gen Chiaki, Daisuke Toyouchi, Ryoki Matsukoba, Kazutaka Kimura, and Yosuke Enomoto for the fruitful discussions and comments.
We sincerely appreciate Volker Springel for the development of the simulation code {\tt GADGET-3} we make use of, which is essential for our calculations.
The numerical simulations were carried out on XC50  {\tt Aterui II} at the Center for Computational Astrophysics (CfCA) of the National Astronomical Observatory of Japan.
This research could never be accomplished without the support by Grants-in-Aid for Scientific Research (TH:19H01934, 21H00041) from the Japan Society for the Promotion of Science and JST SPRING, Grant Number JPMJSP2110.
We use the SPH visualization tool {\tt SPLASH} \citep{DanielPrice2007, DanielPrice2011} in Figs \ref{fig:splash_cosmological}, \ref{fig:splash_Penetrate}, \ref{fig:splash_Mdot}, and \ref{fig:splash_others}.

\section*{DATA AVAILABILITY}
The data underlying this article will be shared on reasonable request to the corresponding author.



\bibliographystyle{mnras}
\bibliography{ms} 



\appendix

\section{Evaluating shock positions}
\label{sec:shockposition}

We evaluate the shock front position of the accretion flows by the following method.
\begin{itemize}
\item[(1)] We assume that the unshocked gas should have infalling velocity larger than the virial velocity 
($v_{\rm inf}>v_{\rm vir}\equiv \sqrt{GM_{\rm halo}/r_{\rm vir}}$)
and extract them as the component of the cold accretion. 
\item[(2)] 
Next we evaluate the accretion rate of unshocked gas $\dot{M}_{\rm fast}$ as a function of the distance $r$ from the halo centre, 
which is defined as
\begin{equation}
\dot{M}_{\text{fast}}(r) = v_{\rm inf}\frac{{\rm d}m_{\text{fast}}(<r)}{{\rm d}r}
=\frac{1}{\delta r}\left(\sum_{ \ \ r\leq r_i<r+\delta r} m_i \  v_{{\rm inf},i}\right),
\end{equation}
where $m_\text{fast}(<r)$ is the enclosed mass of the fast component within the distance $r$ from the halo centre and
the last summation runs over the gas particle of fast components and 
$m_i$ and $v_{\text{inf},i}$ are the mass and the infall velocity of the gas particle.
We also obtain total mass accretion rate $\dot{M}_{\rm tot}$ shown in Figs.~\ref{fig:splash_Penetrate}~ and ~\ref{fig:r-Mdot} by running summation over all gas particles.
\item[(3)] We define the shock radius, at which $\dot{M}_\text{fast}$ decreases below a critical value, $\dot{M}_\text{crit}$.
As the fast accretion flow penetrates inside the halo virial radius,
some of the gas experiences shock heating,
that reduces the amount of cold and fast accretion flow.
Our simulation shows that $\dot{M}_{\rm fast} (r)$ decreases with the decreasing $r$,
indicating that the shock heating actually reduces the amount of the cold accretion flow.
We define the critical accretion rate as
$\dot{M}_\text{crit} \equiv 0.1\times \dot{M}_{\rm typ}$,
where $\dot{M}_{\rm typ}$ is the typical mass accretion rate 
expected for the halo at a given redshift,
\begin{eqnarray}
  \dot{M}_{\rm typ}&\equiv& \frac{M_{\rm halo, gas}}{t_{\rm ff}}
  = 3\pi f_{\rm br}\Omega_{\rm M,0}^{1/2}H_0^2 M_{\rm halo}(1+z)^{3/2}
\end{eqnarray}
where $f_{\rm br}=\Omega_{\rm br}/(\Omega_{\rm DM}+\Omega_{\rm br})$ and the matter density in halo is given as $\rho_{\rm vir}=18\pi^2 \bar{\rho}_0 (1+z_{\rm vir})^3$. 
Since $\dot{M}_\text{fast}$ sharply decreases at some radius, 
the shock radius is insensitive to the choice of the factor before $\dot{M}_\text{typ}$.
\end{itemize}


The above analysis is performed for the snapshots with the time interval of $\Delta t \simeq 2.3 \ {\rm Myr} \{(1+z)/15\}^{-3/2}$, 
which is much shorter than the gas accretion timescale $t_{\rm ff}\sim t_{\rm Hubble}\simeq 0.30 \ {\rm Gyr} \ \{(1+z)/15\}^{-3/2}$. 
This allows us to fully capture how the fast accretion flow penetrates into the halo centre.

Our analysis is different from those adopted in the previous studies, 
where the shock position of the cold accretion is determined by the increase of the entropy \citep{WiseAbel2007,Fernandez2014}. 
The reason is that, in the case of an ACH, it is difficult to separate 
whether the gas in cold accretion heated by adiabatic compression or 
the gas heated by virial shock undergoes Ly$\alpha$ cooling,
since both cases reproduce the similar density and temperature and thus similar entropy.
In previous studies, only one snapshot for each halo has been analysed
and they do not focus on the quantitative evaluation of the shock position and its time evolution,
which is adequate for treating the early stage of cold accretion, before the accretion flow reaches the halo centre.

\bsp	
\label{lastpage}
\end{document}